\documentclass[twocolumn,superscriptaddress, notitlepage,amssymb, nobibnotes,nofootinbib,longbibliography]{revtex4-1}
\usepackage{subfiles}
\usepackage{dcolumn,bm,color,float,amsmath,amssymb,graphicx,braket}
\usepackage[colorlinks=true, allcolors=blue]{hyperref}

\allowdisplaybreaks

\begin{document}

\title{Lifetime-limited Gigahertz-frequency Mechanical Oscillators\\ with Millisecond Coherence Times}

\author{Yizhi~Luo}
\email{yizhi.luo@yale.edu}
\affiliation{Department of Applied Physics, Yale University, New Haven, CT 06520, USA}

\author{Hilel~Hagai~Diamandi}
\affiliation{Department of Applied Physics, Yale University, New Haven, CT 06520, USA}

\author{Hanshi~Li}
\affiliation{Department of Material Science, Yale University, New Haven, CT 06520, USA}

\author{Runjiang~Bi}
\affiliation{Department of Applied Physics, Yale University, New Haven, CT 06520, USA}

\author{David~Mason}
\affiliation{Department of Applied Physics, Yale University, New Haven, CT 06520, USA}

\author{Taekwan~Yoon}
\affiliation{Department of Applied Physics, Yale University, New Haven, CT 06520, USA}

\author{Xinghan~Guo}
\affiliation{Department of Applied Physics, Yale University, New Haven, CT 06520, USA}

\author{Hanlin~Tang}
\affiliation{Department of Applied Physics, Yale University, New Haven, CT 06520, USA}

\author{Ryan~O.~Behunin}
\affiliation{Department of Applied Physics, Northern Arizona University, Flagstaff, AZ 86011, USA}
\affiliation{Center for Materials Interfaces in Research and Applications, Flagstaff, AZ 86011, USA}

\author{Frederick~J.~Walker}
\affiliation{Department of Applied Physics, Yale University, New Haven, CT 06520, USA}

\author{Charles~Ahn}
\affiliation{Department of Applied Physics, Yale University, New Haven, CT 06520, USA}
\affiliation{Department of Material Science, Yale University, New Haven, CT 06520, USA}

\author{Peter~T.~Rakich}
\email{peter.rakich@yale.edu}
\affiliation{Department of Applied Physics, Yale University, New Haven, CT 06520, USA}

\date{\today}

\begin{abstract}
    \noindent \textbf{Abstract:}
    High-frequency mechanical oscillators with long coherence times are essential to realizing a variety of high-fidelity quantum sensors, transducers, and memories. 
    However, the unprecedented coherence times needed for quantum applications require exquisitely sensitive new techniques to probe the material origins of phonon decoherence and new strategies to mitigate decoherence in mechanical oscillators.
    Here, we combine non-invasive laser spectroscopy techniques with materials analysis to identify key sources of phonon decoherence in crystalline media. 
    Using micro-fabricated high-overtone bulk acoustic-wave resonators (\textmu HBARs) as an experimental testbed, we identify phonon-surface interactions as the dominant source of phonon decoherence in crystalline quartz; lattice distortion, subsurface damage, and high concentration of elemental impurities near the crystal surface are identified as the likely causes.
    Removal of this compromised surface layer using an optimized polishing process is seen to greatly enhance coherence times, enabling \textmu HBARs with Q-factors of $>240$ million at 12 GHz frequencies, corresponding to $> 6$~ms phonon coherence times and record-level $f\!-\!Q$~products.
    Complementary phonon linewidth and time-domain ring-down measurements, performed using a new Brillouin-based pump-probe spectroscopy technique, reveal negligible dephasing within these oscillators. Building on these results, we identify a path to $>100$~ms coherence times as the basis for high-frequency quantum memories.
    These findings clearly demonstrate that, with enhanced control over surfaces, dissipation and noise can be significantly reduced in a wide range of quantum systems.
    
\end{abstract}

\maketitle

\section{Introduction}\vspace{-10pt}

Long-lived phonons are a compelling resource, as they permit numerous quantum operations within their coherence time, enabling high-performance quantum sensors \cite{goryachev2014gravitational,lo2016acoustic,carney2021,linehan2024listening}, transducers \cite{jiang2020efficient,mirhosseini2020superconducting,yoon2023,simon2024}, and memories \cite{hann2019hardware,simon2020,bozkurt2024mechanical}. 
Efficient control of long-lived phonons using optomechanical \cite{safavi2012observation,kharel2018,painter2020}, electromechanical \cite{bourhill_generation_2020,mohammad2023quantum}, and superconducting qubit systems \cite{satzinger2018quantum,chu2018creation,amir2022,chu2022} has generated renewed interest in phononic device physics and technologies for quantum applications~\cite{delsing20192019,barzanjeh2022optomechanics}.
While a diversity of mechanical oscillators has produced such long-lived phonons over a range of frequencies \cite{aspelmeyer2014,barzanjeh2022optomechanics}, high-frequency (gigahertz) phonons are often desirable, as they have improved immunity to unwanted noise, permit ground-state operation at cryogenic temperatures, and are more readily controlled using quantum optics and circuit-QED techniques.
In theory, crystalline media are ideal for hosting such long-lived phonons, as they have vanishing internal dissipation at cryogenic temperatures \cite{liekens1971on,liekens1971attenuation,tamura1985spontaneous,scheffold1997sound}.
However, it has proven difficult to extend the coherence times of such gigahertz-frequency crystalline oscillators to millisecond timescales. 

Silicon-based nanomechanical phononic crystal resonators have shown long phonon lifetimes ($>1$ second); however, strong phonon-TLS coupling in these systems limits their coherence times to $\sim\!100$~\textmu s~\cite{painter2020,bozkurt2024mechanical,simon2020}. 
This is because the tight phonon confinement and strong boundary reflections within these systems make them vulnerable to complex surface interactions that introduce excess noise and dephasing \cite{painter2020}. 
Alternatively, micro-fabricated high-overtone bulk acoustic wave resonators (\textmu HBARs) of the type seen in Fig.~\ref{fig:fig1}a  produce phonon modes with orders of magnitude lower surface participation \cite{kharel2018,renninger2018}.
In principle, such \textmu HBAR may offer lower dephasing rates, translating to much longer coherence times.  However, in practice, they have yielded modest coherence times ($<1$~ms) \cite{kharel2018, chu2018creation}, shorter than can be explained by device geometry, suggesting a material origin. 

A variety of complex interactions can contribute to phonon dissipation and decoherence in such systems.
While internal damping due to anharmonic phonon-phonon scattering becomes vanishingly small at low temperatures~\cite{liekens1971on,liekens1971attenuation,tamura1985spontaneous}, scattering by impurities, dislocations, and lattice distortions produce additional loss mechanisms within the bulk of a crystal~\cite{srivastava2022physics,cleland2013foundations}.
Surfaces and material interfaces bring a variety of additional decoherence and loss mechanisms~\cite{klitsner1987phonon}.
Processes for cutting, polishing~\cite{cui2023review}, and etching~\cite{strunk1988damage} of crystal surfaces introduce lattice distortions, dislocations, subsurface damage, and elemental impurities (Fig.~\ref{fig:fig1}c) that can contribute to excess phonon scattering~\cite{klitsner1987phonon} as well as complex defect-phonon interactions~\cite{vanelstraete1990tunneling}.
Moreover, surface roughness can produce excess dissipation through radiative losses~\cite{galliou2013}.
Hence, improved oscillator performance will require sensitive new techniques to probe the material origins of phonon decoherence and new strategies to extend phonon coherence times.

Here, we combine new non-invasive laser spectroscopy techniques with materials analysis to identify the origins of phonon dissipation and decoherence in crystalline media. Informed by these studies, we demonstrate new device designs and fabrication techniques that enable quartz \textmu HBARs with Q-factors as high as 247 million at 12.66~GHz, corresponding to a record-level $f-Q$ product of $3.13 \times 10^{18}$~Hz and a phonon coherence time of 6.2~ms. Complementary spectral linewidth and coherent ring-down measurements, performed using a new Brillouin-based pump-probe spectroscopy technique, reveal negligible dephasing within these oscillators. To investigate the mechanisms of phonon decoherence, the bulk and surface contributions are analyzed by varying the geometry of the \textmu HBAR.
These studies indicate that surfaces are the dominant source of phonon decoherence, with surface losses far exceeding those predicted from roughness-induced boundary scattering.
Lattice distortions, subsurface damage, and high concentrations of elemental impurities are identified as likely sources of excess surface loss using advanced materials characterization techniques.
Removal of this compromised layer using an optimized polishing process yields a 10-fold reduction in surface loss, enabling the realization of \textmu HBARs with the record-level phonon coherence times described above as a compelling resource for circuit-QED and cavity-optomechanical systems \cite{chu2018creation,chu2022,diamandi2024quantum} (see Fig.~\ref{fig:fig1}d-e). 
This result clearly demonstrates that enhanced control over surfaces can translate to dramatic improvements in dissipation and noise, with potential implications for a variety of quantum systems.
\vspace{-10pt}

\section{Contributions to Phonon Decoherence}

\begin{figure}[t!]
\centering\includegraphics[width=1.0\linewidth]{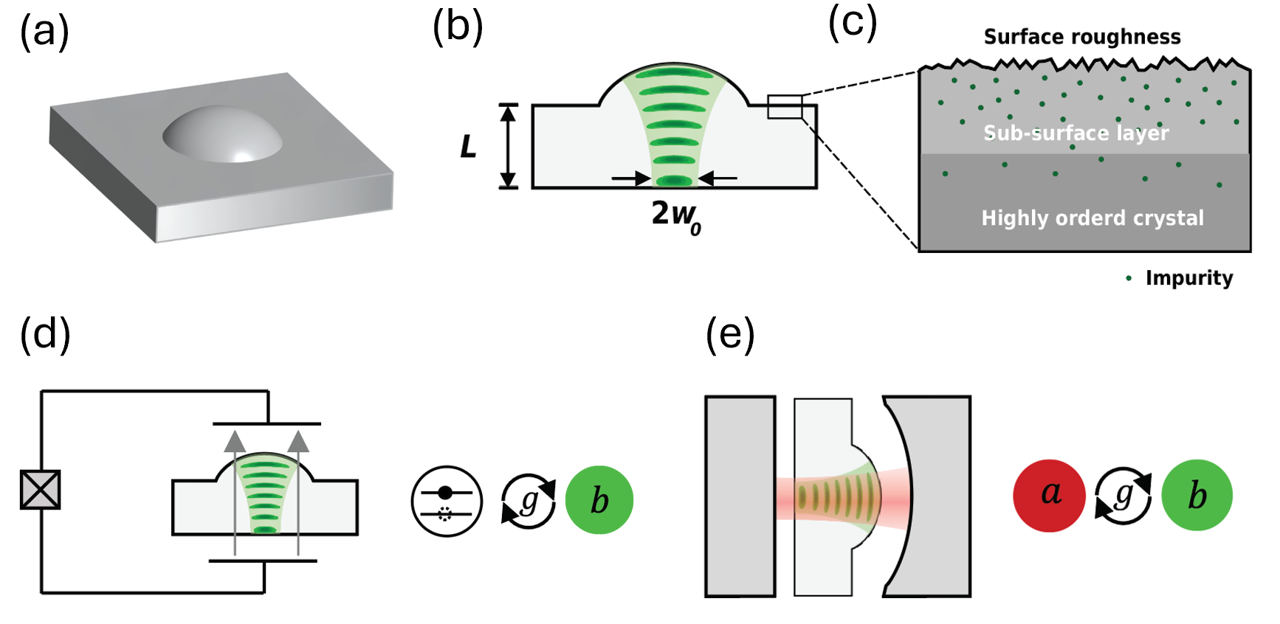}
    \caption{
    \textbf{\textmu HBAR device and applications:} Sketch of 3D device gometry (a) and cross-sectional view of \textmu HBAR device (b) that supports a stable Gaussian mode with a waist radius, $w_0$, determined by the device radius of curvature, $R$, cavity length, $L$. 
    Magnified view of \textmu HBAR surface (c) showing surface roughness, subsurface damages, and impurity contamination as possible sources of phonon dissipation.
    (d) Schematic of \textmu HBAR coupling to superconducting qubit.
    (e) Schematic of \textmu HBAR coupling to optical cavity.
    }
    \label{fig:fig1}
\end{figure}

We investigate phonon dissipation and decoherence in crystalline quartz using a \textmu HBAR device of the type seen in Fig.~\ref{fig:fig1}a.
These \textmu HBARs are created by shaping the surfaces of a quartz substrate into a plano-convex geometry.
Within the crystal, reflections from these shaped surfaces effectively produce a stable plano-concave Fabry-Perot resonator for longitudinal bulk acoustic waves, supporting a series of high-Q-factor Gaussian modes~\cite{kharel2018}.
The linewidth and lifetime of the longitudinal phonon modes supported by these resonators are measured using a new laser-based spectroscopy technique to study the coherence properties of these \textmu HBARs (see Section III and Methods). 

The spectral linewidth of such phonon modes is determined by both loss and dephasing, with contributions arising from bulk and surface interactions.
Using bulk ($\Gamma^\textup{bulk}_\textup{loss}$) and surface loss rates ($\Gamma^\textup{surf}_\textup{loss}$) to capture the effect of these numerous microscopic loss mechanisms, the total phonon energy loss rate ($\Gamma_\textup{loss}$) becomes $\Gamma_\textup{loss}=\Gamma^\textup{bulk}_\textup{loss}+ \Gamma^\textup{surf}_\textup{loss}$. Note that $\Gamma^\textup{bulk}_\textup{loss}$ is independent of \textmu HBAR cavity length $L$, whereas $\Gamma^\textup{surf}_\textup{loss}$ scales as $1/L$. Assuming a fractional energy loss per surface reflection of $l_\textup{surf}\ll 1$ for both plano and concave resonator surfaces, the surface loss rate becomes $\Gamma_\textup{surf} = 2\,l_\textup{surf}\times \textup{FSR} = \,l_\textup{surf} \,(v_a/L)$, where $\textup{FSR}=v_a/2L$ is the acoustic free spectral range and $v_a$ is the longitudinal acoustic velocity.
Fluctuations in the mechanical properties of the elastic medium also introduce a dephasing rate ($\Gamma_\phi$) that produces broadening of measured phonon linewidth ($\Delta\Omega$) as $\Delta\Omega=\Gamma_\textup{loss} + \Gamma_\phi $. 
Note that $\Gamma_\phi$ can be decomposed into distinct bulk ($\smash{\Gamma^\textup{bulk}_\phi}$) and surface ($\smash{\Gamma^\textup{surf}_\phi}$) components, each of which contributes to linewidth broadening. 
In general, the phonon coherence time ($\tau_\textup{coh}$) and Q-factor ($Q$) are calculated from the measured Lorentzian linewidth as $\tau_\textup{coh}=2/{\Delta\Omega}$ and $Q= \Omega/\Delta \Omega$, respectively~\cite{goodman2015statistical}. Hence, the phonon coherence time can be directly obtained from the measured linewidth or Q-factor. (See SI section VII for details.)

A crucial step in understanding phonon decoherence is to determine whether surface or bulk interactions limit coherence times.
In this context, cavity finesse, defined as $F = 2\pi\times\textup{FSR}/\Delta\Omega$, is another useful quantity, indicating the number of round trips (or surface interactions) within the coherence time.
In the limit where surface interactions predominantly limit phonon coherence times, the linewidth becomes $\Delta\Omega= \Gamma^\textup{surf}_\textup{loss}+ \Gamma^\textup{surf}_{\phi}$. In this case, the Q-factor increases linearly with $L$, and $F$ is independent of $L$ (see SI section VII).
In the special case when dephasing is also negligible, the finesse takes the form $F = \pi\, l_\textup{surf}^{-1}$, and the Q-factor becomes $Q= \Omega/\Gamma_\textup{surf} = \frac{L}{l_\textup{surf}}\frac{2\pi}{\lambda_{ph}}$, where $\lambda_\textup{ph}$ is the acoustic wavelength.

\begin{figure*}[t!]
    \centering
    \includegraphics[width=1.0\linewidth]{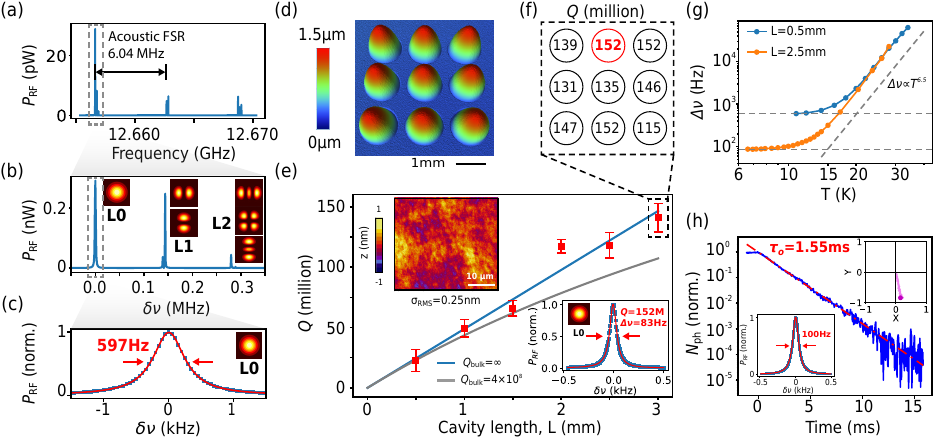}
    \caption{
    \textbf{Phonon spectroscopy of \textmu HBAR arrays:} 
    Panels (a) - (c) show the measured phonon spectra of a $L=0.5$~mm \textmu HBAR device. Broad spectral scan in (a) shows a longitudinal mode family that repeats with an acoustic free spectral range (FSR) of 6.04 MHz. 
    Higher resolution scans in panels (b) and (c) show the transverse Hermite-Gaussian (HG) modes within a mode family, and the spectrum of the fundamental Gaussian mode, respectively. 
    The Lorentzian fit (red) in (c) corresponds to a linewidth of $\Delta\nu=\Delta\Omega/2\pi=597$~Hz and a Q-factor of 22 million.
    (d) profile of $L=2.5$~mm \textmu HBAR array measured with an optical profilometer.
    (e) Plot showing measured phonon Q-factor versus cavity length; each data point (square) indicates the average measured Q-factor of 6 different $3\times 3$ \textmu HBAR arrays fabricated on substrates with thicknesses (cavity lengths) ranging from $L=0.5$~mm to $L=3.0$~mm. Error bars indicate the variance of the measured Q-factor distribution.
    The linear fit (blue) is consistent with a constant finesse of 12,224, corresponding to an average energy loss per surface reflection of $l_\textup{surf}=257$~ppm assuming a bulk phonon loss ($\Gamma^\textup{bulk}_\textup{loss}$) of zero or $Q_\textup{bulk} =\Omega/\Gamma^\textup{bulk}_\textup{loss}=\infty$. 
    For comparison, the grey trend line, which shows predicted Q-factors in the case of $Q_\textup{bulk} = 400$~million, overestimates the bulk phonon losses. 
    Inset top left: high-resolution surface topography image (80 \textmu m by 60 \textmu m) of the dome center of a \textmu HBAR by an optical profilometer, showing RMS roughness of 2.5 \AA. 
    Inset bottom right: Phonon spectrum of an individual \textmu HBAR in panel (f). Lorentzian fit (red) yields the linewidth $\Delta\nu=\Delta\Omega/2\pi=83$~Hz, corresponding to a Q-factor of 152~million and a coherence time of 3.84~ms.
    (f) Map of measured Q-factors within the $L=3.0$~mm \textmu HBAR array in panel (d).
    (g) Temperature dependence of phonon linewidth for both $L=0.5$ and $L=2.5$~mm \textmu HBARs. For $T>18$~K, the phonon linewidth follows predicted $T^{6.5}$ temperature dependence (dashed grey) due to bulk phonon-phonon scattering; for $T<10$ K, the $L= 0.5$~mm \textmu HBAR linewidth plateaus at $\Delta\nu = 597$~Hz, and the 2.5~mm \textmu HBAR linewidth plateaus at $\Delta\nu=86$~Hz. 
    (h) Phonon ringdown  $L=2.5$~mm \textmu HBAR. Exponential fitting yields a lifetime of $\tau_\textup{o}= 1.55$~ms, corresponding to a Q-factor of 124~million and a energy dissipation rate $\Gamma_\textup{loss}/2\pi=103$~Hz. Inset top right: Coherent phonon ringdown plotted in a quadrature plane. Inset bottom left: phonon spectrum of this \textmu HBAR, shows a linewidth $\Delta\nu=\Delta\Omega/2\pi=100$~Hz.
    }
    \label{fig:fig2_3}
\end{figure*}

In what follows, we examine the surface and bulk contributions to phonon decoherence by systematically varying the \textmu HBARs  cavity length, $L$, and by extension, the surface contribution to phonon decoherence. 
Arrays of \textmu HBARs, seen in Fig.~\ref{fig:fig1}a, are fabricated from vendor polished z-cut $\alpha$-quartz (optical grade) using a reflow-based fabrication technique \cite{kharel2018}.
The fabricated \textmu HBARs have a radius of curvature, $R= 100$~mm, which produces longitudinal acoustic modes with waist radii ($w_0$) ranging from $30-50$~\textmu m at 12~GHz phonon frequencies ($\Omega$) for the cavity lengths ($L$) used in this study.
The fabricated \textmu HBARs have a surface roughness ($\sigma$) of $\sim\!2.5$~\AA ~on both plano and concave surfaces (measured by white-light optical profiler with cutoff spatial frequency of $2\pi/500\,\textup{nm}^{-1}$, see top left inset of Fig.~\ref{fig:fig2_3}e), limiting the roughness-induced scattering loss to  $l_\textup{RMS}\cong (4\pi\sigma/\lambda_\textup{ph})^2 \cong 40\times 10^{-6}$ or $40$~parts per million (ppm) on each surface \cite{galliou2013}.
Since the 1~mm clear aperture of the concave resonator surface is more than 10 times larger than the spot size, clipping losses are negligible ($\ll 1$~ppm).
Hence, in the absence of dephasing and material loss, an \textmu HBAR with $L=3$~mm should support scattering-limited finesse of $80,000$, corresponding to Q-factors (coherence time) as high as 1~billion (25~ms) at 12~GHz frequencies.\vspace{-10pt}

\section{Pump-Probe Spectroscopy of \textmu HBAR Phonons}

We probe the coherence of these \textmu HBAR phonon modes using a novel non-invasive laser-based pump-probe spectroscopy technique that utilizes phase-matched acousto-optic coupling to access \textmu HBAR phonons near the Brillouin frequency ($\Omega_\textup{B}/2\pi \cong 12.66$~GHz).
This non-invasive spectroscopy method employs spectrally distinct pump and probe waves to transduce and detect phonons in the \textmu HBAR using a standing-wave configuration (Methods, Fig.~\ref{fig:fig4}a).
The improved phase stability offered by this standing-wave configuration has the advantage of eliminating the need for active interferometric phase stabilization of prior methods \cite{renninger2018}, while enabling robust high-resolution (sub-Hertz) spectral measurements.
This technique also enables complementary coherent phonon ringdown measurements that permit studies of phonon dephasing in \textmu HBARs (for details, see Methods).

Fig.~\ref{fig:fig2_3}a-c show the phonon spectra from a \textmu HBAR with $L= 0.5$ mm taken at 12.66 GHz frequencies at a temperature of $T= 10$~K. The broad spectral scan of Fig.~\ref{fig:fig2_3}a shows three sets of resonances that repeat every 6.04 MHz, corresponding to the FSR of longitudinal acoustic modes. A magnified view of the first set of resonances is seen in Fig.~\ref{fig:fig2_3}b, which corresponds to the fundamental Gaussian mode and higher-order Hermite-Gaussian (HG) modes, labeled as L0, L1, and L2. Frequency splittings among L1 and L2 are attributed to slight asymmetry in the shape of reflowed resist during the fabrication process. A high-resolution scan of the L0 resonance, seen in Fig.~\ref{fig:fig2_3}c, reveals a phonon linewidth of $\Delta\Omega/2\pi = 597$ Hz, corresponding to a Q-factor of 22~million, and to a cavity finesse of 11,000.

This measured Q-factor is seven times lower than the scattering-loss-limited Q-factor ($\sim160$~million) predicted based on roughness measurements, suggesting the presence of additional sources of decoherence from the surfaces or the bulk of the crystal.
Complementary measurements of linewidth versus temperature (Fig.~\ref{fig:fig2_3}g, blue) reveal a clear reduction in internal damping, consistent with the $T^{6.5}$ temperature dependence predicted by Landau-Rumer theory of phonon-phonon scattering in quartz \cite{liekens1971on,scheffold1997sound}.
For $T$ below 10~K, the phonon linewidth plateaus at $\Delta \Omega/ 2 \pi =597$~Hz, indicating that temperature-independent loss mechanisms (other than Landau-Rumer) are dominant at the base temperatures.

To determine whether the dominant loss mechanism is produced by surface or bulk interactions, we perform similar measurements on arrays of \textmu HBARs having a range of cavity lengths ($L=0.5,1.0, 1.5,2.0, 2.5,3.0$~mm).
To obtain statistically meaningful dependence of the phonon Q-factor on cavity length ($L$), we averaged the measured Q-factors from a $3\times 3$ array of \textmu HBARs (see Fig.~\ref{fig:fig2_3}d) for each cavity length, yielding the plot in Fig.~\ref{fig:fig2_3}e. These data show a clear increase in Q-factor with cavity length.  The mean Q-factor increases from $23\pm 9$ million for $L = 0.5$~mm to a maximum value of $141 \pm 12$ million for $L = 3.0$~mm. Fig.~\ref{fig:fig2_3}f shows a map of the measured Q-factors for the $L=3.0$~mm \textmu HBAR array. 
As seen in Fig.~\ref{fig:fig2_3}f and the bottom right inset of Fig.~\ref{fig:fig2_3}e, an individual resonator reaches a Q-factor (coherence time) as high as 152 million (3.84 ms), and temperature-dependent Q-factor measurements using an $L=2.5$~mm \textmu HBAR (Fig.~\ref{fig:fig2_3}g, orange) again show a negligible loss contribution from Landau-Rumer scattering.
The data in Fig.~\ref{fig:fig2_3}e show good agreement with a linear fit (blue), corresponding to the case of a constant cavity finesse of 12,224 and $\Gamma^\textup{bulk}_\textup{loss}=0$.
Hence, these data indicate that the phonon linewidths and coherence times are primarily limited by surface interactions, such that the phonon linewidth takes the form $\Delta \Omega = \Gamma^\textup{surf}_\textup{loss} + \Gamma^{\textup{surf}}_\phi$.

Ringdown measurements are performed to determine the relative contributions of the surface-induced loss ($\Gamma^\textup{surf}_\textup{loss}$) and dephasing rate ($\Gamma^{\textup{surf}}_\phi$).
By abruptly switching off the pump-wave excitation while simultaneously measuring the phonon amplitude with the probe wave, we measure free-induction decay of the excited \textmu HBAR phonon modes (see methods).
Fig.~\ref{fig:fig2_3}h shows a typical single-shot ringdown measurement performed using the fundamental (L0) mode of a 2.5~mm \textmu HBAR for comparison with a back-to-back linewidth measurement (inset bottom left).
The temporal trace (blue) shows the normalized phonon occupation following the abrupt turn-off of the pump tones at $t = 0$; complementary phase trajectory data from the recorded ringdown are shown in the upper right inset. 
An exponential fit of the energy decay trace reveals a 1.55~ms energy decay time ($\tau_\textup{o}$), corresponding to $\Gamma_\textup{loss}/2\pi = (2\pi\cdotp \tau_\textup{o})^{-1} = 103$~Hz. Since this energy decay rate ($\Gamma_\textup{loss}$) shows excellent agreement with the measured phonon linewidth ($\Delta \Omega/ 2 \pi = 100$~Hz) in Fig.~\ref{fig:fig2_3}h (bottom left inset), we conclude that dephasing has a negligible contribution to the phonon decoherence (i.e., $\Gamma^\textup{surf}_\textup{loss} \gg \Gamma^{\textup{surf}}_\phi$) in this system. Hence, the fitted finesse of 12,224 indicates that these \textmu HBARs are surface-loss limited, with an average fractional energy loss per reflection of $l_\textup{surf} \cong 257$~ppm.
These measured surface losses far exceed those predicted from roughness-induced boundary scattering ($l_\textup{RMS} \cong 40$~ppm), suggesting sources of excess scattering loss near the crystal surface.\vspace{-10pt}

\section{Sources of Dissipation and Techniques to Enhance Phonon Coherence}

Next, we investigate the material origins of excess surface losses and explore strategies to reduce them. Likely sources of this excess surface loss include impurity scattering and structural disorder of the crystal lattice near the crystal surface. To investigate reactive-ion etching (RIE) as a possible source of excess surface loss, we exposed fabricated \textmu HBARs to an additional RIE etch. These tests revealed that reactive ion etching results in excess surface losses. An 80 minute RIE exposure was seen to introduce $\sim\!320$~ppm of excess round-trip surface loss (see SI section IV for details). 
Such losses could be explained by ion implantation and damage to the crystal lattice resulting from RIE exposure~\cite{strunk1988damage}.

\begin{figure*}[t!]
    \centering    
    \includegraphics[width=1\linewidth]{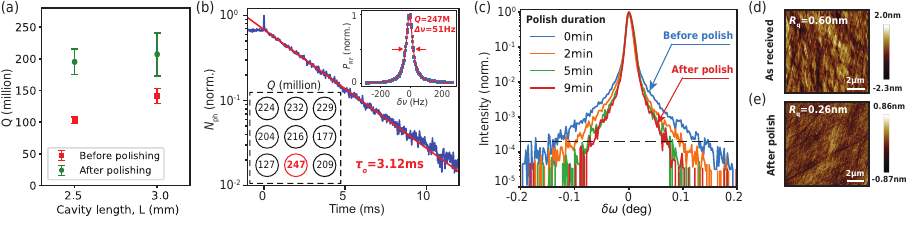}
    \caption{\textbf{Improved phonon coherence after optimized polishing:}  
    (a) Q-factor statistics of \textmu HBAR devices on 2.5~mm and 3.0~mm substrates before (red) and after (green) repolishing on their backside surfaces. Before repolishing, the Q-factor statistics are $103\pm 5$ million for 2.5~mm substrate and $141\pm 12$ million for 3.0~mm substrate; after repolishing, the Q-factor statistics are $195\pm 20$ million for 2.5~mm substrate and $209\pm 37$ million for 3.0~mm substrate.  
    (b) Phonon ringdown of a \textmu HBAR device on the repolished 3 mm substrate, showing a phonon lifetime of 3.12~ms, which corresponds to an energy dissipation rate of $\Gamma_\textup{loss}/2\pi = 51$~Hz. Top right inset shows the measured phonon spectrum of the same device, showing a linewidth of 51~Hz, which matches with the energy dissipation rate. Bottom left inset shows the measured spectral Q-factors of all 9 \textmu HBAR devices on this 3 mm substrate. 
    (c) X-ray rocking curve measurement of subsurface lattice misalignment after polishing for 0, 2, 5, 9 min. We use FW5000M (black dashed line) to quantify the angular width of the rocking curves. The FW5000M values are 0.260, 0.188, 0.146, 0.126 deg respectively. As we polish the surface, native subsurface damages are being removed, resulting in a more regularly ordered subsurface lattice structure, and the rocking curve eventually converges to a narrower peak. 
    (d) AFM image of plano substrate surface as received. A large amount of deep native polishing lines can be clearly seen, and the RMS roughness is measured to be 0.60~nm in an area of 10 by 10 \textmu m. 
    (e) AFM image of plano substrate surface after polish. Most native polishing lines are removed and the remaining ones are much shallower. The RMS roughness after polish is 0.26~nm.}
    \label{fig:fig3}
\end{figure*}

To investigate RIE-induced ion implantation as a possible source of excess loss, we analyze the elemental constituents near the surface of the crystal using secondary ion mass spectroscopy (SIMS) as a function of depth beneath the crystal surface.
These measurements reveal significant concentrations of elemental impurities such as C, Al, F, Na, and Fe (see SI section VI). While the unetched polished surface has significantly higher Al, Na, and Fe concentrations, the etched surface shows a higher concentration of F. Both etched and unetched surfaces show similar amounts of C, but with a slightly different depth profile. 
These results suggest that while RIE helps eliminate residual impurities (Al, Na, Fe) by removing contaminated material beneath the crystal surface, the high-energy plasma in the RIE etcher also introduces excess impurities from the SF$_6$ and Ar gases used during the etch process (See SI Fig.~S4).

To avoid contamination and subsurface damage introduced by the RIE process, we intentionally minimize the overetch time during the fabrication process.
The central region of the curved \textmu HBAR surface, which interacts most strongly with the phonon mode, receives less than $10$~minutes of RIE exposure during device fabrication, corresponding to the removal of $<\!500$~nm of material.
Assuming a linear relationship between RIE exposure and induced surface losses, this fabrication process leads us to expect $<\!40$~ppm of excess surface loss due to the RIE etch.
Hence, RIE exposure is unlikely to account for more than 10\% of the total loss, meaning that the surface losses for the plano and concave surfaces of the fabricated \textmu HBAR are approximately equal.

Several approaches to remediation of the crystal surfaces were attempted. These included etching, annealing, and polishing of crystal surfaces (see SI section V).
However, of these, only repolishing of the crystal surface using a carefully optimized process produced a significant improvement in phonon coherence times.
Glancing incidence X-ray diffraction (GIXRD) measurements were used in conjunction with high-resolution AFM measurements to assess lattice distortions and subsurface damage during optimization of the polishing process. Through GIXRD, the crystal surface is illuminated with X-rays at a small glancing angle ($\sim\!1$~deg) such that the X-rays penetrate only a small distance ($\sim\! 1$ \textmu m) below the crystal surface; this ensures that the X-rays only probe the crystal lattice near the crystal surface \cite{letts2018x}. The angular width of the X-ray diffraction order, measured from a rocking curve, is then used to quantify lattice distortion and disorder. Example GIXRD measurements taken during the polishing process are seen in Fig.~\ref{fig:fig3}c. Before the surface is polished, the rocking curve associated with the $[0,-2,3]$ Bragg plane shows a 0.260~deg angular width, measured as the full width at 1/5000th maximum (FW5000M). Following 9 minutes of polishing with a 40~nm silica suspension, this angular width becomes reduced to 0.126~deg, indicating that the X-rays are sampling a more regularly ordered lattice near the crystal surface.   

High-resolution surface profile measurements obtained using AFM are also used as an indicator of lattice distortions and subsurface damage. The degree of strain, distortion, and damage beneath the crystal surface are strongly correlated with the depth of any polishing lines visible from high-resolution AFM imaging \cite{lee2018evaluation,carr1999subsurface}.
While such nano-scale scratches introduce negligible surface-scattering loss due to their high spatial frequencies, the corresponding lattice distortions beneath the surface extend over a much larger volume via plastic deformation (dislocations) of the crystal lattice, and thus can have a much more significant impact on intracavity phonons. 
Hence, the metrics used to optimize the polishing process are the RMS roughness, the depth and density of any polishing lines, and the angular width of X-ray rocking curve, which has been associated with scattering from lattice distortions \cite{letts2018x,larson1974huang}. Example AFM images of the natively polished surface (Fig.~\ref{fig:fig3}d) are shown alongside the repolished surface (Fig.~\ref{fig:fig3}e). The latter reveals much less pronounced polishing lines with an RMS surface roughness of $R_\textup{q}\cong 2.6$\AA. 
Note that these AFM-based RMS roughness measurements have significant high spatial frequency components (i.e, $\gg 2\pi / 500~\textup{nm}^{-1}$) that do not contribute to phonon scattering loss. 
RMS roughness measurements with frequency components more relevant to phonon scattering (obtained by the while-light optical profiler) show negligible change through the repolishing process. 

To investigate the impact of this surface remediation method, we measure the coherence times of resonator arrays before and after repolishing the planar \textmu HBAR surface.
Note that distinct \textmu HBAR arrays, with performance comparable to those seen in Fig.~\ref{fig:fig2_3}, were used for this surface remediation study. 
The measured Q-factors of $L=2.5$~mm and $L=3.0$~mm \textmu HBAR devices, before (red) and after (green) repolishing are seen in Fig.~\ref{fig:fig3}a.
Both $2.5$~mm and $3.0$~mm \textmu HBARs show significant improvement in the average Q factor after repolishing. 
The average Q-factor of the $2.5$~mm \textmu HBAR increases from $103\pm 5$~million to $197\pm 20$~million, and the average Q-factor of the $3.0$~mm \textmu HBAR increases from $141\pm 12$~million to $209\pm 37$~million.
Note that the nearly two-fold reduction in total loss observed for the $2.5$~mm \textmu HBAR indicates a drastic decrease in surface loss.
Assuming approximately equal losses for plano and concave surfaces, the increase in Q-factor for the $2.5$~mm ($3.0$~mm) device is consistent with a surface loss reduction from 305~ppm (268~ppm) to 17~ppm (93~ppm) for the plano surface, corresponding to a 94\% (65\%) reduction in surface loss from the repolished surface.
Hence, these results corroborate the hypothesis that subsurface damage is indeed the dominant source of loss. 
It is also interesting to note that deviations in the measured \textmu HBAR Q-factors from the constant finesse trend line of Fig.~\ref{fig:fig2_3}e are readily explained by variations in the level of subsurface damage produced through the polishing process. 

This surface remediation method also produces resonators with record-level phonon coherence times and $f-Q$ products. 
As seen from the inset of Fig.~\ref{fig:fig3}b, seven resonators within the 3~mm \textmu HBAR array produce Q-factors $\geq 200$~million.
As seen from the spectral measurement of Fig.~\ref{fig:fig3}b, an individual resonator on the repolished 3.0~mm device reaches a Q-factor as high as 247 million, corresponding to a linewidth of $\Delta \Omega/2\pi = 51.3$~Hz.
Since the measured lineshape is Lorentzian, it follows that coherence time is given by $\tau_\textup{coh}=2/\Delta \Omega=6.2$~ms.
Complementary ringdown measurements seen in Fig.~\ref{fig:fig3}b reveal an energy-decay time of $\tau_\textup{o} = 3.12$~ms and a corresponding energy decay rate of $\Gamma_\textup{loss}/2\pi = (2\pi\cdotp \tau_\textup{o})^{-1} = 55.6$~Hz.
Close agreement between the measured energy decay rate ($\Gamma_\textup{loss}$) and phonon linewidths ($\Delta \Omega$) reveal that dephasing has a negligible contribution to phonon decoherence. These Q-factors and coherence times correspond to a record-level $f$-$\,Q $ product of $3.13\times 10^{18}$~Hz, a quantity that provides a measure of an oscillator's coherence and its immunity to thermal decoherence \cite{aspelmeyer2014}.

\section{Discussion and Conclusions}

Building on these results, quartz shows the potential to support gigahertz-frequency phonons with coherence times far exceeding those achieved here.
Through these studies, we have identified subsurface damage and impurities at crystal surfaces as the dominant source of phonon decoherence in crystalline quartz.  
The removal of this compromised surface layer using an optimized polishing process yielded a 10-fold reduction in surface losses, corroborating this hypothesis and demonstrating the potential to greatly enhance phonon coherence times with improved control of crystal surfaces and interfaces.
Hence, in the absence of surface losses, $\alpha$-quartz can likely support Q-factors exceeding 1~billion at 12~GHz frequencies.
Applying such optimized surface treatments to both plano and concave surfaces, coherence times of $\tau_\textup{coh}>30$~ms and finesse levels of $F> 100,000$ are likely achievable, benefiting a variety of applications.

\textmu HBARs devices of the type demonstrated here can find direct applications in both cavity optomechanical and circuit QED systems. 
Cavity-optomechanical techniques have recently been used to achieve laser cooling of such quartz \textmu HBARs to their ground state (Fig.~\ref{fig:fig1}e), paving the way for quantum optomechanical control of such long-lived phonon modes~\cite{diamandi2024quantum}. These techniques also enable the realization of high optomechanical coupling rates ($\sim 14$~MHz), which are essential for high-speed operations~\cite{kharel2019high, kharel2022multimode}. Building on these results, optomechanical control of such highly coherent \textmu HBARs opens new applications in areas such as quantum transduction~\cite{yoon2023}, networking~\cite{simon2020}, and computing~\cite{maggie2024}.

Piezoelectric coupling \cite{galliou2013,chu2018creation} of such \textmu HBARs to superconducting qubits (Fig.~\ref{fig:fig1}d) also enables advanced forms of quantum state synthesis and tomography \cite{chu2018creation,chu2022}.
At lower phonon frequencies ($\leq\! 5$~GHz) typically used in superconducting qubit studies, µHBARs could achieve much longer coherence times.
Longer oscillation periods at these lower frequencies combined with the $1/\lambda^2_\textup{ph}$ scaling of scattering losses~\cite{galliou2013} translate to a 6-fold increase in phonon coherence times, corresponding to $\tau_\textup{coh} > 40$~ms in the current \textmu HBARs.  Moreover, remediation of both \textmu HBAR surfaces could translate to much longer ($\tau_\textup{coh} > 150$~ms) coherence times, making such systems a compelling resource for quantum random access memories~\cite{hann2019hardware}. 

Looking ahead, mastery of phonon-surface interactions becomes even more critical as one shrinks the phononic device size, since increasing surface participation makes them more sensitive to surface imperfections. 
Hence, optimized surface treatments that reduce surface-induced decoherence will likely translate to even more dramatic improvements within micro- and nano-mechanical systems.
Since the methods for studying phonon decoherence and surface-phonon interactions demonstrated here can be applied to a wide range of crystals~\cite{kharel2018,kharel2019utilizing}, they could enable exploration of diverse material platforms for hybrid quantum systems.
The methods for systematic examination of phonon decoherence and surface-phonon interactions demonstrated here can be applied to a wide range of crystals~\cite{kharel2018,kharel2019utilizing}, enabling exploration of diverse material platforms for hybrid quantum systems.

\begin{center}
  \rule{0.3\textwidth}{0.8pt}\vspace{-11.3pt} \rule{0.2\textwidth}{1.2pt} 
\end{center}

\section*{Methods}

\begin{figure}[h]
    \centering
    \includegraphics[width=1\linewidth]{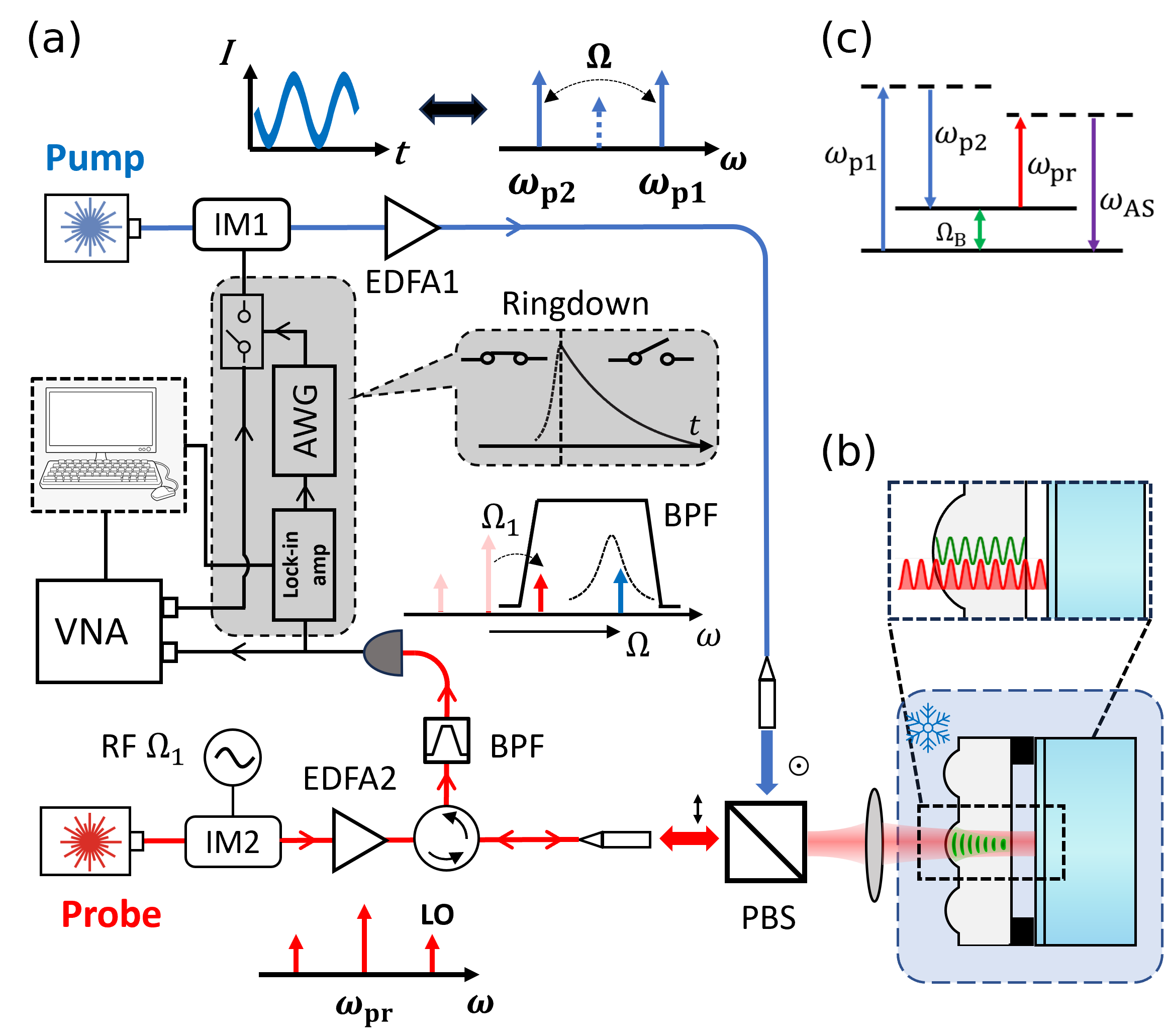}
    \caption{
    (a) Schematic of CABS setup.
    The pump laser (blue) is centered at 1549.068~nm, s-polarized, passing through by an intensity modulator (IM1) driven by a vector network analyzer (VNA) at $\sim$6.33~GHz, at null-bias configuration. 
    The probe laser (red) is centered at 1549.120~nm, p-polarized. A local oscillator (LO) signal is imprinted on the probe light by another intensity modulator IM2, driven at 12.651~GHz. Both pump and probe lights then combine at a polarization beamsplitter (PBS) and shine into our \textmu HBAR device and enable both Stokes and anti-Stokes scattering of the probe light through acousto-optic interaction. Then the anti-Stokes light (p-polarized, same as the probe light), together with the LO, is transmitted through a band-pass filter (BPF) and detected by a photodetector.
    The beat note output from the photodetector, usually at $\sim\!10$~MHz, is then split in half; one goes to the receiving port of the VNA for spectral measurements, while the other one goes to an arbitrary waveform generator (AWG) for ringdown measurements. 
    The \textmu HBAR device is mounted on an optical mirror with a spacer in between and cooled down to 6~K. 
    (b) The laser intensity pattern (red) needs to match with the phonon displacement pattern (green) to satisfy the phase matching condition. 
    (c) Energy diagram of the pump-probe CABS process.
    }
    \label{fig:fig4}
\end{figure}

The measurements in Sections~III-IV were obtained using the two-color pump-probe spectroscopy technique, presented here for the first time. Fig.~\ref{fig:fig4}a shows the reflection-mode apparatus used to perform non-invasive Brillouin based measurements of the fabricated \textmu HBARs. Pump (blue) and probe (red) waves having distinct wavelengths are used to simultaneously transduce and detect phonons within the \textmu HBAR, enabling background-free measurement of the transduced elastic wave motion. The \textmu HBAR array is mounted on a dielectric mirror such that both the pump and probe waves produce standing-wave field patterns, closely matching those of the standing-wave phonon modes within the \textmu HBAR (Fig.~\ref{fig:fig4}b). This permits efficient Brillouin (or acousto-optic) coupling to the \textmu HBAR phonon modes while eliminating the need for active interferometric stabilization required using prior methods \cite{renninger2018, kharel2018}. 
The pump and probe waves are focused to a spot size closely matching the fundamental Gaussian modes of \textmu HBAR, which permits efficient coupling to phonon modes with frequencies near the Brillouin frequency of z-cut quartz ($\Omega_\textup{B} /2\pi \cong 12.66$~GHz) using 1550 nm light.  (see SI section IX for further details)

The frequencies of the pump- and probe-waves are chosen to fall within the Brillouin phase-matching bandwidth, enabling efficient transduction and detection of elastic wave motion using the coherent anti-Stokes Brillouin scattering (CABS) process, diagrammed in Fig.~\ref{fig:fig4}c.
Phonons are first generated within the \textmu HBAR using an intensity-modulated pump wave (blue) that creates two optical tones with a frequency separation ($\Omega$) near the Brillouin frequency ($\Omega_\textup{B}$), as seen in the upper arm of Fig. ~\ref{fig:fig4}a. 
Stimulated Stokes scattering produces energy transfer between these optical pump tones, exciting phonons within the \textmu HBAR. 
The elastic wave motion associated with these phonons is simultaneously detected using a continuous-wave probe laser (red). 
This elastic wave motion imprints Stokes and anti-Stokes sidebands on the probe wave through phase-matched Brillouin scattering (lower arm of Fig. ~\ref{fig:fig4}a). Heterodyne detection of the anti-Stokes sideband is then used to measure phase and amplitude of the coherently driven phonons.

To prevent optical cross-talk, orthogonally polarized pump (s-polarized) and probe (p-polarized) waves are used to excite and detect the phonons within the \textmu HBAR resonator. The pump wave (s-polarized) is synthesized by modulating 1549.068~nm laser light using an intensity modulator (IM1) that is driven by a vector network analyzer (VNA). This modulator (IM1) is operated at the null-bias point to produce two 1st-order sideband tones with frequency separation, $\Omega$, when the intensity modulator is driven at frequency $\Omega/2$. This permits excitation of the \textmu HBAR phonon modes when the frequency separation ($\Omega$) is tuned through the Brillouin frequency ($\Omega_\textup{B}$).  This pump wave is then amplified with an EDFA to boost its optical intensity before entering the PBS to excite phonons within the \textmu HBAR. 
A 1549.120~nm probe wave (p-polarized) is amplified using a second EDFA before passing through the PBS to interact with the \textmu HBAR.
The phase and amplitude of the excited phonons are measured through heterodyne detection of an anti-Stokes sideband that is imprinted on the probe wave through Brillouin scattering.

Coherent measurement of the elastic wave motion is performed by detecting the heterodyne beat note produced by the interference of the anti-Stokes sideband with an optical local oscillator (LO).
The optical LO tone is imprinted on the probe-wave using a separate intensity modulator (IM2), driven at frequency $\Omega_1$. While intensity modulation produces both $+\Omega_1$ and $-\Omega_1$ sidebands, only the $+\Omega_1$ is used as the optical LO. The reflected probe is then bandpass-filtered such that only anti-Stokes sideband and the $+\Omega_1$ LO tone are transmitted, enabling coherent detection of the elastic-wave motion at frequency $\Omega-\Omega_1$ using a high-speed photodetector. Since the detected RF signal power is proportional to the anti-Stokes optical power, the RF power has a linear proportionality with to the phonon population inside the \textmu HBAR device. 

During frequency domain spectral measurements, the VNA is used to sweep the drive frequency ($\Omega$) through the \textmu HBAR resonance while simultaneously measuring the phase and amplitude of the anti-Stokes beat note at frequency $\Omega-\Omega_1$.
A slow sweep speed is used to ensure a steady-state measurement of the resonant response.
During ringdown measurements, as the VNA sweeps the drive frequency through the cavity resonance, the intensity modulation of the pump-wave is abruptly switched off when the drive frequency reaches the peak resonance of the target phonon mode ($\Omega_\textup{o}$). The lock-in amplifier then demodulates the anti-Stokes beat note at $\Omega_\textup{o}-\Omega_1$ and records its phase and amplitude during free-induction decay. 
This abrupt turn-off is enabled by adding a fast RF switch between the VNA output and IM1, which is controlled by an arbitrary waveform generator (AWG).
The abrupt increase in amplitude of the anti-Stokes beat note during resonant excitation of the \textmu HBAR is used to trigger the AWG, which then causes the RF switch to open, abruptly turn off the RF drive to IM1, and permit lock-in measurement of free-induction decay of the \textmu HBAR phonon mode.

\begin{center}
  \rule{0.3\textwidth}{0.8pt}\vspace{-11.3pt} \rule{0.2\textwidth}{1.2pt} 
\end{center}

\section*{Acknowledgments}
All \textmu HBAR devices are fabricated in Yale university cleanroom. We thank Dr. Yong Sun and Dr. Lauren McCabe for assistance with device fabrication. We thank Dr. John Vig and Dr. Serge Galliou for helpful discussions of crystalline quartz and treatment methods. Primary support for this research was provided by the U.S. Department of Energy (DoE), Office of Science, National Quantum Information Science Research Centers, Co-design Center for Quantum Advantage (C2QA) under contract No. DE-SC0012704. We also acknowledge supported by the Air Force Office of Scientific Research (AFOSR) and the Office of Naval Research under award No. FA9550-23-1-0338 and the National Science Foundation (NSF) under TAQS award No. 2137740 and QLCI Award No. OMA - 2016244. HHD acknowledges support from the Fulbright Israel program. Any opinions, findings, and conclusions or recommendations expressed in this material are those of the author(s) and do not necessarily reflect the views of the DoE, AFOSR, or NSF.

\section*{Data availability}
The data supporting the findings of this study are available from the authors upon reasonable request.

\section*{Author Contribution} 
PTR, YL conceived and planned the experiments. YL and RB carried out the device fabrication, optical experiments and data analysis with the assistance of HHD, DM, and TY. YL did the material characterization with the help of RB, HT, HL, and XG. HL did the polishing. ROB, FW, CA contributed to the interpretation of the results. YL and PTR wrote the manuscript with inputs from all the authors. PTR supervised this work.

\section*{Competing Interests} 
P.T.R is a founder and shareholder of Resonance Micro Technologies Inc. 

\clearpage


%

\clearpage
\setcounter{equation}{0}
\setcounter{figure}{0}
\setcounter{table}{0}
\setcounter{page}{1}
\setcounter{section}{0}
\renewcommand{\theequation}{S\arabic{equation}}
\renewcommand{\thefigure}{S\arabic{figure}}
\renewcommand{\thetable}{S\arabic{table}}

\title{Supplementary Information}
\maketitle
\onecolumngrid 

\section{Reflow-based fabrication method} \label{SI_reflow}

\begin{figure}[h]
\centering
\includegraphics[width=0.8\linewidth]{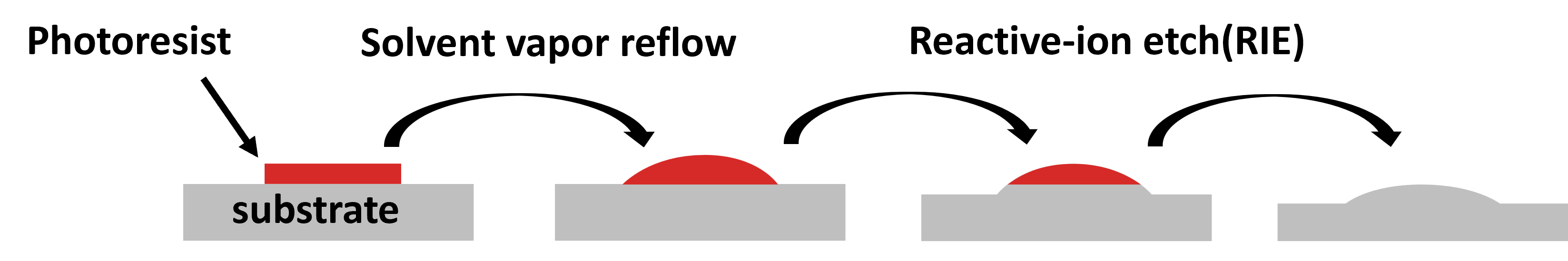}
\caption{Flow chart of the \textmu HBAR fabrication steps.}
\label{fig:reflow}
\end{figure}

Here we describe the fabrication procedure for quartz \textmu HBARs, which is based on this paper \cite{kharel2018} but with the latest updates. As one can see from Fig.~\ref{fig:reflow}, the whole fabrication recipe can be divided into three big steps: (1) making photoresist cylinders; (2) solvent vapor reflow; (3) reactive ion etching (RIE). 

1. Making photoresist disk.

Fabrication begins with a cleaning procedure that involves organic solvent sonication, Piranha cleaning, and plasma ashing. Soak a double-side-polished z-cut quartz substrate in NMP, acetone, methanol, respectively, and sonicate for 3~min each. Then dip it into Piranha solution (H2SO4:H2O2=3:1) for 8~min. After Piranha cleaning, the substrate is oxygen plasma cleaned for 1~min at a RF power of 100~W and a pressure of 300~mTorr. This cleaning procedure is for removing any organic contaminants and adsorbates from the substrate surfaces. We then pre-bake it at 120 $^{\circ}$C for 5~min to get rid of adsorbed water molecules, spin coat a 1.5-\textmu m-thick layer of photoresist (S1808), and post-bake it at 120 $^{\circ}$C for another 2~min to harden the photoresist. Standard UV lithography by a laser writer (Heidelberg MLA 150) is then applied to define the cylinder pattern (an array of 1-mm-diameter circles). After UV exposure, the substrate is then developed in MF-319 solution and the photoresist cylinder pattern is formed. 

2. Solvent vapor reflow

Next, we do the solvent vapor reflow to turn the cylinder to a dome shape. Before the reflow, the photoresist cylinders need to be primed with the resist adhesion promoter hexamethyldisilizane (HMDS) vapor for 15~min to preserve the circular contact boundary. Then the substrate is heated at $\sim 60 ^{\circ}$C and placed upside down on top of a beaker half-filled with the reflow solvent (not touching the substrate), polypropylene glycol monomethyl ether acetate (PGMEA). The beaker is heated up to $\sim 57 ^{\circ}$C to vaporize and the vapor is absorbed into the photoresist. After absorbing the PGMEA vapor, the photoresist becomes softened and liquid-like. Under the effect of surface tension and gravity, the photoresist slowly changes its shape from cylinder to a dome. Note that the reflow time is highly sensitive to the temperature difference between the substrate and the solvent vapor, thus it has to be carefully optimized before formal fabrication. Once the dome shape is formed, we stop the reflow and hard-bake the photoresist. The hard baking procedure is 90 $^{\circ}$C for 1~min, 110 $^{\circ}$C for 5~min, and gradually up to 130 $^{\circ}$C for another 5~min. After hard baking, the substrate is now ready for the last step, i.e. reactive ion etching. 

3. Reactive ion etching (RIE)

The purpose of RIE is to replicate the dome shape of the reflowed photoresist onto the substrate. A slow RIE is implemented in Plasmalab 80+ (Oxford Instrument) using SF6 and Ar gases with flow rates of 4 sccm and 14 sccm respectively, at a chamber pressure of 4.5 mTorr and a bias voltage of 440~V to completely remove the photoresist. With a combination of both chemical and physical etch, quartz material is removed at $\sim 50$ nm/min and the photoresist is removed at $\sim 100$ nm/min. This yields a substrate surface with excellent roughness ($\sim 2.5$\AA), which is critical to minimize phonon scattering loss and realize high-Q \textmu HBARs. 

Note that by changing the diameter and thickness of the photoresist pattern, we can largely tune the radius of curvature (from 10~\textmu m to 1~m) of the dome shape after RIE. Moreover, this method is not constrained to any specific material, making it a versatile and powerful technique to make \textmu HBAR devices. 

\section{Sample mount and the "windshield"} \label{SI_windshield}

\begin{figure}[h]
\centering
\includegraphics[width=0.4\linewidth]{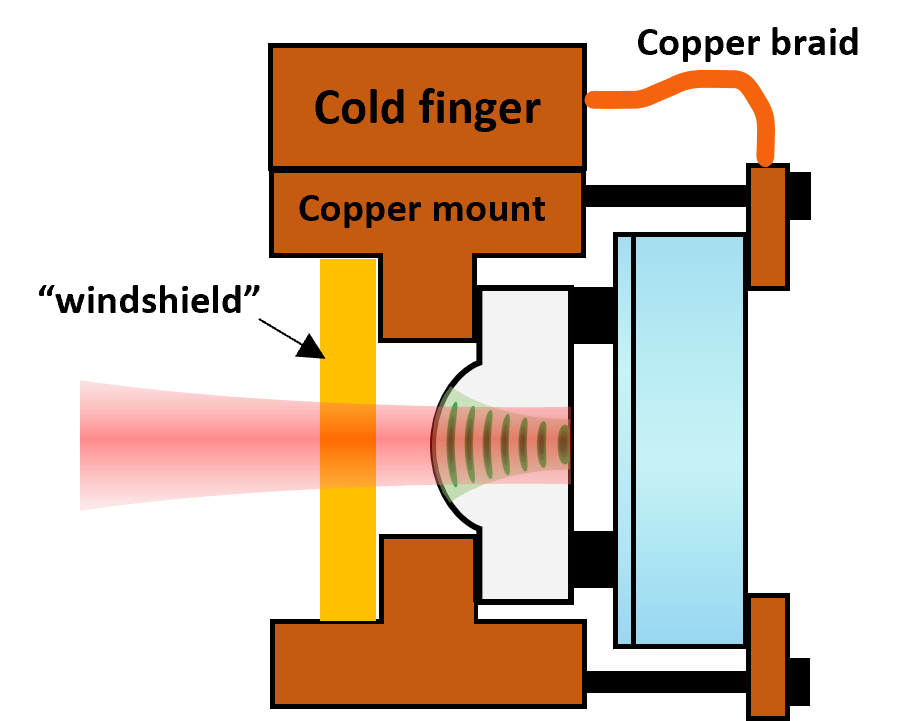}
\caption{Illustration of the sample mount with a cold ``windshield" that is used to protect the \textmu HBAR array from adsorbates.}
\label{fig:windshield}
\end{figure}

As mentioned in the main text, all phonon measurements in this work were performed at liquid helium temperature ($\sim 4$~K). The sample device is installed inside a Janis ST-100 cryostat and cooled either by a recirculation gas cooler (RGC4) or a helium dewar (for better temperature stability). We machined an oxygen-free high-conductivity (OFHC) copper to match the size of the \textmu HBAR samples, which is directly connected to the cold finger of the cryostat. The sample is sandwiched between the copper mount and an optical mirror. Then the optical mirror is clamped from the other side by a copper plate via spring-loaded screws. A spacer is placed between the sample and the mirror to avoid phonon leakage into the mirror. The sample sits on the copper mount and the copper clamp is connected to the cold finger with a copper braid for optimal cooling. 

To obtain consistent and repeatable \textmu HBAR linewidth measurements, it was necessary to design the sample environment to minimize the adsorption of residual gas molecules on the \textmu HBAR surface. 
Prior to cooling down, the cryostat is pumped for at least 3 hours to reach a pressure below 1e-5 hPa. 
Since air molecules have a near-unity adsorption probability when they make contact with cold material surfaces (i.e., temperatures below 100~Kelvin) and have a very low probability of desorption, this is sometimes referred to as the ``hit-and-stick" regime of adsorption dynamics.
Hence, adsorption of molecules to the \textmu HBAR surface can be prevented by surrounding the crystal surfaces by cold objects and ensuring that there is no ballistic trajectory that permit molecules in the cryostat chamber to attach the \textmu HBAR surface.
To prevent residual gasses within the cryostat from condensing onto the sample, the plano and convex surfaces of \textmu HBAR are protected using the sample holder in Fig.~\ref{fig:windshield}.
Attachment of the \textmu HBAR to the mirror offers protection to the plano surface. However, it was also important to introduce a cold ``windshield" to prevent the convex \textmu HBAR surface from being directly exposed to cryostat chamber (Fig.~\ref{fig:windshield}). 
Through these experimetns, an AR-coated glass window is used, however any transparent substrate can be used.
Without the windshield, we observed a rapid degradation in the \textmu HBAR Q-factor and a time-dependent red-shift of the phonon modes. 
However, with the introduction of the cold windshield, these problems were eliminated, permitting stable and repeatable measurements of phonon Q-factor without any drift of phonon frequency.

\section{Cleaning procedure before phonon measurements} \label{SI_cleaning}
Since the phonon dissipation is limited by the surface interaction, our \textmu HBAR devices need to be carefully cleaned before put in the cryostat. We followed a standard cleaning protocol before each cryogenic measurement: i) sonication with organic solvent NMP, acetone, methanol, 3 min each; ii) Piranha cleaning for 8 min, and rinse it for 10 min with DI water; iii) (optional) oxygen plasma ashing for 1 min at a RF power of 100 W and a pressure of 300 mTorr. After the cleaning procedures, the sample has to be transferred and installed inside the cryostat as soon as possible, because exposure to the air environment increases the amount of water adsorbates at the surface. Putting it in a vacuum transfer box is generally a good practice.

\section{RIE-introduced phonon losses} \label{SI_RIE_losses}

\begin{figure}[h]
\centering
\includegraphics[width=0.9\linewidth]{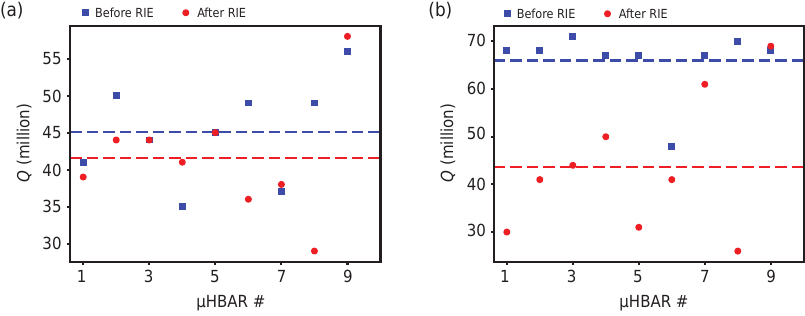}
\caption{
(a) Backside (plano surface) over-etch test of substrate CQT15\_1 that mimics the etch on the front side (dome surface). Before etching, the average Q of all nine \textmu HBARs is $45\pm 6$ million; after etching, the average Q of all nine \textmu HBARs is $41\pm 7$ million.
(b) 80 min over-etch test of substrate CQT15\_8 (40 min on each surface). Before etch, the average Q of all nine \textmu HBARs is $66\pm 6$ million; after etch, the average Q of all nine \textmu HBARs is $43\pm 13$ million.
}
\label{fig:RIEcompare}
\end{figure}

The RIE process is one of the most critical processes in the fabrication of \textmu HBAR devices. It uses chemically reactive plasma to remove material from the substrate surface. In this process, high-energy ions are generated and accelerated before they hit the substrate. Constant bombardment with these high-energy ions unavoidably causes damage and contamination in the subsurface layer within the penetration depth. In order to imprint the dome shape on the substrate surface while minimizing the potential damage and contamination in the subsurface, we intentionally minimize the over-etch duration ($<10$ min) in the central dome area with which the phonon field interacts the most. We did a similar RIE test on the backside of a 1.5 mm substrate (CQT15\_1) that mimics the etch experience of the front side. As you can see from Fig.~\ref{fig:RIEcompare}a, the measured Q-factors of 8 out of 9 \textmu HBARs don't show significant variance from before the backside etch. On average, the Q-factors are $45\pm 6$ million before and $41\pm 7$ million after the etch. 
However, if we intentionally apply an extra 40 min etch on both surfaces (80 min in total) of another \textmu HBAR substrate (CQT15\_8, see Fig.~\ref{fig:RIEcompare}b), significant Q degradation is observed. In this case, the average Q-factor drops from $66\pm 6$ million before to $43\pm 13$ million after etch. Long-time RIE over-etch does introduce extra phonon losses. Therefore, minimizing the over-etch duration is critical to achieving high-Q \textmu HBAR devices. 

\section{Quartz substrate repolishing} \label{SI_repolish}
All the \textmu HBARs presented in this work are fabricated on z-cut quartz substrates. As we demonstrated in this study, the phonon loss in our devices is dominated by surface/subsurface interaction. Although commercial quartz substrates usually come with decent surface roughness ($<5$\AA), it is well-known that they possess a damaged layer under the surface due to complex lapping and polishing histories \cite{lee2018evaluation,carr1999subsurface}. Therefore, an optimized repolishing is critical to achieve record-level Q-factors. In this work, a slurry repolishing with colloidal fused silica suspension is applied to the plano surface of \textmu HBAR devices to minimize newly-introduced surface damages. This polishing suspension features 40~nm grit size and an aqueous solution with pH of 10. The polishing is done with a Multiprep$^\textup{TM}$ polishing machine with Red Final C adhesive back disc from Allied High Tech Products Inc. The polishing procedure takes 12~min, during which we keep adding fresh suspension every 3~min. 
We characterized the polished surface using atomic force microscopy (AFM) and glancing-angle X-ray rocking curve (GIXRC) method. 
As the material is gradually removed from the surface, the rocking curve peak converges to a narrower linewidth, suggesting a more regularly ordered lattice structure after polishing. On the other hand, AFM results also show that finer polishing with 40 nm fused silica grits helps remove the native polish lines on the surfaces of commercial quartz substrates, yielding a much smoother surface. 

\section{Secondary-ion mass spectroscopy (SIMS)} \label{SI_sims}

\begin{figure}[h]
\centering
\includegraphics[width=0.5\linewidth]{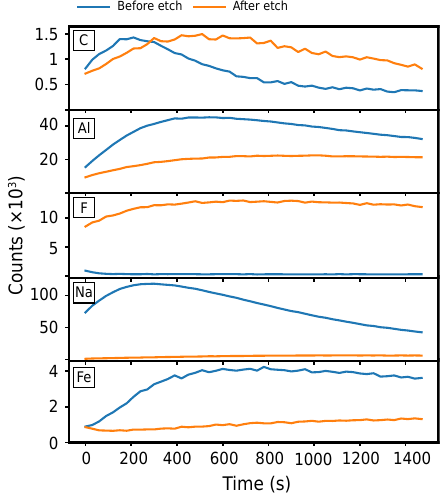}
\caption{SIMS results before and after RIE process.}
\label{fig:sims}
\end{figure}

We use SIMS to study ion implantation during the RIE process. Fig.~\ref{fig:sims} shows the depth profiling of impurity concentrations for two 2 mm substrates, one blank, and one fabricated \textmu HBAR device. For the device, SIMS is done on the dome side, i.e. the RIE-processed side. A significant amount of impurity elements such as C, Al, F, Na, Fe are observed. As you can see from Fig.~\ref{fig:sims}, the RIE-processed device shows decreased concentrations of Al, Na, and Fe. These impurities are imprinted into the substrate surface by native grinding and polishing. RIE, by removing material with the high-energy plasma, reduces the amount of native impurities. On the other hand, the high-energy plasma, containing mostly SF6 and Ar, implants impurity elements into the substrate surface as well. This is why the concentration of F is much higher in the device than the blank substrate. At last, C shows almost the same level of concentration but a different distribution, because C exists in the native substrate and can be introduced by the high-energy plasma. Therefore the RIE process didn't reduce its total concentration but changed its distribution instead.

\section{Lifetime and coherence time} \label{SI_coherencetime}
\begin{figure}
    \centering
    \includegraphics[width=0.5\linewidth]{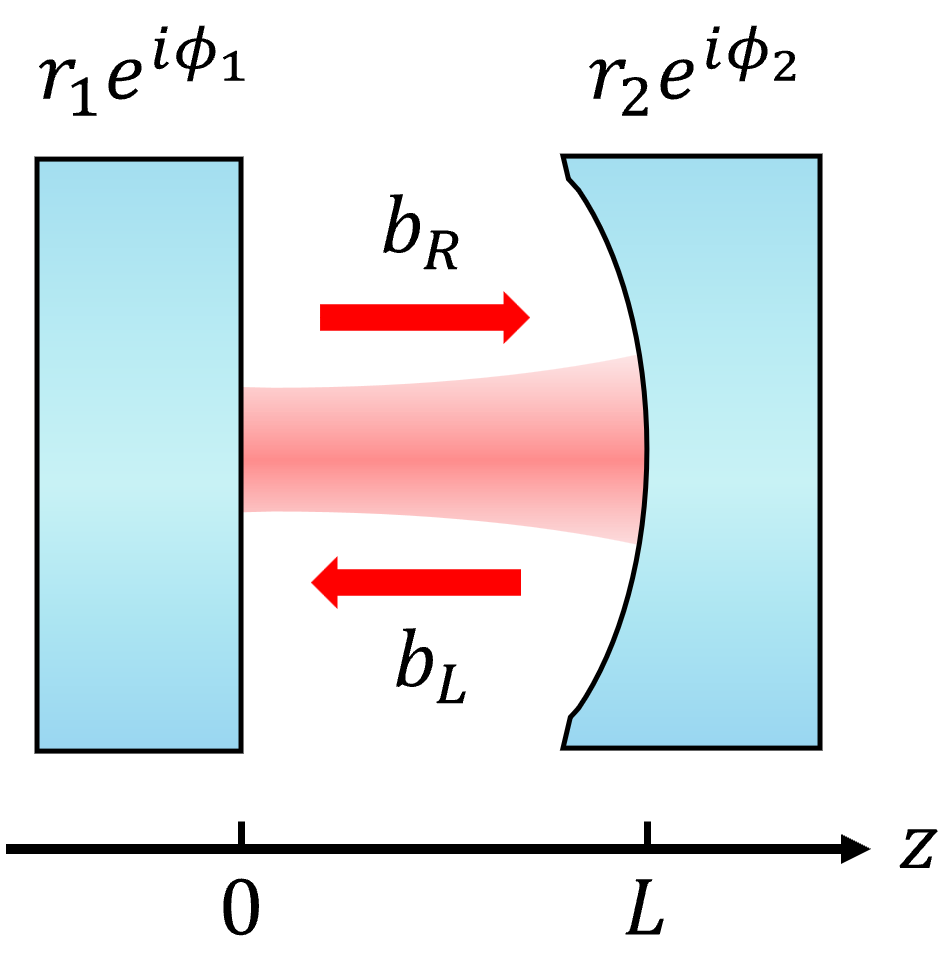}
    \caption{Schematic of a wave field propagating inside a FP cavity of length $L$.}
    \label{fig:Ch4_cavity}
\end{figure}
Consider a FP cavity of length $L$ along z-direction where a field (optical or acoustic) with wave vector $k$ propagates back and forth (see Fig.~\ref{fig:Ch4_cavity}). The left and right moving fields are represented by $b_L(z,t)$ and $b_R(z,t)$, respectively. 
Assume the absorption coefficient is $\alpha$ and the reflection coefficients of the two mirrors are $r_1e^{\phi_1}$ and $r_2e^{\phi_2}$ where $r_1,r_2<1$ and $\phi_1$ ($\phi_2$) is a random variable describing the phase noise added by the reflection of mirror 1 (2). 
Quite straightforwardly, we have
\begin{equation} \label{prop}
    \begin{split}
        b_R(L,t) &= b_R(0,t-T/2)e^{ikL-\alpha L/2},\\
        b_L(0,t) &= b_L(L,t-T/2)e^{ikL-\alpha L/2},\\
    \end{split}
\end{equation}
and the boundary condition for the mirror reflection gives 
\begin{equation} \label{refl}
    \begin{split}
        b_R(0,t) &= b_L(0,t)\cdot r_1e^{i\phi_1(t)},\\
        b_L(L,t) &= b_R(L,t)\cdot r_2e^{i\phi_2(t)},\\
    \end{split}
\end{equation}
where $T$ is the round-trip propagation time, i.e. $T=2L/v$. 
Therefore, after one round of propagation, the field at $z=L$ can be expressed by 
\begin{equation} \label{recur}
\begin{split}
    b_R(L,t) &= r_1 r_2 e^{i(\phi_1(t-T/2)+\phi_2(t-T))} e^{-\alpha L+2ikL}\cdot b(L,t-T)\\
    &= R e^{i\phi(t)}\cdot b_R(L,t-T).\\
\end{split}
\end{equation}
Here $r_1$ and $r_2$ represent the surface loss, $\alpha$ represents the bulk loss, $R=r_1 r_2 e^{-\alpha L+2ikL}$, $\phi(t)=\phi_1(t-T/2)+\phi_2(t-T)$ is the stochastic phase noise, which as you will see later, is the source of pure dephasing. From now on, I will only focus on the dynamics of the field and replace $b_R(z,t)$ with $b(t)$. 
Using Eq.~\ref{recur} as a recursive relation between fields after one round of propagation, we can get 
\begin{equation} \label{phonon_dynamics}
    \begin{split}
        b(t=nT) &= R e^{i\phi(t)}\cdot b(t-T)\\
        &= R^2 e^{i[\phi(t)+\phi(t-T)]}\cdot b(t-2T)\\
        &= ... \\
        &= R^n \exp{[i\sum_{m=0}^{n-1}\phi(t-mT)]}\cdot b(t=0) \quad (n\geq0).\\
    \end{split}
\end{equation}
Therefore, field intensity can be calculated as 
\begin{equation} \label{dissipation}
    \begin{split}
        I(t) &= \langle b^*(t)b(t) \rangle = |R|^{2n} I(0)\\
        &= (r_1 r_2 e^{-\alpha L})^{2n} I(0)\\
        &= I(0)e^{[-\alpha v+\frac{v}{L}\ln{(r_1 r_2)}]t}\\
        &= I(0)e^{-\Gamma_o t} \quad (t\geq0),\\
    \end{split}
\end{equation}
where $v$ is the field propagation velocity and $\Gamma_o=\alpha v-\frac{v}{L}\ln{(r_1 r_2)}$ is the energy dissipation rate. The lifetime is then defined as $\tau_o=1/\Gamma_o$. 
Note that the energy dissipation rate is composed of two parts, the bulk contribution $\alpha L$ and the surface contribution $-\frac{v}{L}l\ln{(r_1 r_2)}$. 

While the stochastic phase noise from the surface doesn't contribute to the phonon dissipation, it makes an important source of dephasing, which is a critical factor especially for quantum applications. 
coherence time $\tau_\text{coh}$ and decoherence rate $\Gamma_\text{decoh}$ are defined to evaluate the dephasing rate of a our phononic resonator. From Eq.~\ref{phonon_dynamics} and Eq.~\ref{dissipation}, we know that the free-induction field can be expressed by 
\begin{equation}
    b(t)=b(0) \exp{[-\Gamma_o t/2 + i\sum_{m=0}^{n-1}\phi(t-mT)]} \quad (t=nT, n\geq0).
\end{equation}
The coherence property of a field is described by the degree of first-order temporal coherence $g^{(1)}(\tau)$, which is essentially the autocorrelation of $b(t)$. Setting time delay as $\tau=l\cdot T$ (note that $\tau$ and $l$ can also be negative), we have 
\begin{equation} \label{g1}
    \begin{split}
        g^{(1)}(\tau) &= \frac{\langle b^*(t) b(t+\tau) \rangle}{\langle b^*(t) b(t) \rangle} \\
        &= \frac{\langle e^{-\Gamma_o t/2} e^{-\Gamma_o (t+\tau)/2} \rangle}{\langle e^{-\Gamma_o t} \rangle} \langle \exp{[-i\sum_{m=0}^{n-1}\phi(t-mT)+i\sum_{m=0}^{n+l-1}\phi(t-mT)]} \rangle,
    \end{split}
\end{equation}
where $\phi(t)=\phi_1(t-T/2)+\phi_2(t-T)$.
The first term in Eq.~\ref{g1} is quite straightforward,
\begin{equation} \label{amp_corr}
    \frac{\langle e^{-\Gamma_o t/2} e^{-\Gamma_o (t+\tau)/2} \rangle}{\langle e^{-\Gamma_o t}\rangle} = e^{-\Gamma_o \tau/2} \frac{\int_{\max{\{0,-\tau\}}}^{+\infty}{e^{-\Gamma_o t} dt}}{\int_{0}^{+\infty}{e^{-\Gamma_o t} dt}} = e^{-\Gamma_o |\tau|/2}.
\end{equation}
The second term in Eq.~\ref{g1} is the autocorrelation of the stochastic phase noise. Assuming the phase noise terms $\phi_1(t)$ and $\phi_2(t)$ to be zero-mean stationary Gaussian random variables and independent on each other, then we have 
\begin{align}
    & \langle \phi_i(t) \rangle = 0, \\
    & \langle \phi_i(t) \phi_i(t') \rangle = C_{\phi_i}(t-t'), \\
    & \langle f_i(\phi_i(t)) \rangle = \langle f_i(\phi_i(t+t_0)) \rangle, \\
    & \langle f_1(\phi_1) f_2(\phi_2)  \rangle = \langle f_1(\phi_1) \rangle \cdot \langle f_2(\phi_2) \rangle, 
\end{align}
where $C_{\phi_i}(t-t')$ is the time autocorrelation function of $\phi_i(t)$ and $f_i(\phi_i)$ is any function of $\phi_i(t)$. 
In the case of $\tau\geq0$, i.e. $l\geq0$, the second term in Eq.~\ref{g1} can be written as 
\begin{equation}
    \begin{split}
        \langle \exp{[-i\sum_{m=0}^{n-1}\phi(t-mT)+i\sum_{m=0}^{n+l-1}\phi(t-mT)]} \rangle &= \langle \exp{[-i\sum_{m=0}^{n-1}\phi_1(t-mT)+i\sum_{m=0}^{n+l-1}\phi_1(t-mT)]} \rangle \cdot \langle \phi_2 \ \text{term} \rangle\\
        &= \langle \exp{[+i\sum_{m=n}^{n+l-1}\phi_1(t-mT)]} \rangle \cdot \langle \phi_2 \ \text{term} \rangle \\
        &= \langle \exp{[+i\sum_{m=0}^{l-1}\phi_1(t-mT)]} \rangle \cdot \langle \phi_2 \ \text{term} \rangle.
    \end{split}
\end{equation}
Since $\phi_1(t)$ obeys the Gaussian probability distribution, so does $\sum_{m=0}^{l-1}\phi_1(t-mT)$. 
Therefore, 
\begin{equation}
    \begin{split}
        \langle \exp{[+i\sum_{m=0}^{l-1}\phi_1(t-mT)]} \rangle &= \exp{[-\frac{1}{2}(\sum_{m=0}^{l-1}\phi_1(t-mT))^2]}\\
        &= \exp{[-\frac{1}{2} \sum_{m,m'=0}^{l-1}\phi_1(t-mT) \phi_1(t-m'T)]}\\
        &= \exp{[-\frac{1}{2} \sum_{m,m'=0}^{l-1}C_{\phi_1}((m-m')T)]}.
    \end{split}
\end{equation}
Assuming the time autocorrelation of the phase noise to be 
\begin{equation}
    C_{\phi_1}(t-t')=\langle \phi_1^2 \rangle e^{-\gamma |m-m'| T/2},
\end{equation}
then we have 
\begin{equation}
    \begin{split}
        \exp{[-\frac{1}{2} \sum_{m,m'=0}^{l-1}C_{\phi_1}((m-m')T)]} &= \exp{[-\frac{\langle \phi_1^2 \rangle}{2} \sum_{m,m'=0}^{l-1}e^{-\gamma |m-m'| T/2}]} \\
        &= \exp{[-\frac{\langle \phi_1^2 \rangle}{2} (l+2\sum_{m=0}^{l-1}\sum_{m'=0}^{m-1}e^{-\gamma (m-m') T/2})]} \\
        &= \exp{[-\frac{\langle \phi_1^2 \rangle}{2} (l-2\sum_{m=0}^{l-1}\frac{1-e^{-m\gamma T/2}}{1-e^{\gamma T/2}})]} \\
        &= \exp{[-\frac{\langle \phi_1^2 \rangle}{2} (l-\frac{l}{2(1-e^{\gamma T/2})}+\frac{1-e^{-l\gamma T/2}}{2(1-e^{\gamma T/2})(1-e^{-\gamma T/2})})]}.
    \end{split}
\end{equation}
If the phase noise can be treated as a Markovian process, i.e. $\gamma T\gg 1$, then its autocorrelation can be simplified to be 
\begin{equation}
    \exp{[-\frac{1}{2} \sum_{m,m'=0}^{l-1}C_{\phi_1}((m-m')T)]} = \exp{[-\frac{l\langle \phi_1^2 \rangle}{2}]},
\end{equation}
and the second term in Eq.\ref{g1} is 
\begin{equation}
    \langle \exp{[-i\sum_{m=0}^{n-1}\phi(t-mT)+i\sum_{m=0}^{n+l-1}\phi(t-mT)]} \rangle = \exp{[-\frac{l}{2} (\langle \phi_1^2 \rangle + \langle \phi_2^2 \rangle)]}.
\end{equation}
Similarly, when $\tau<0$, i.e. $l<0$, we get
\begin{equation}
    \begin{split}
        \langle \exp{[-i\sum_{m=0}^{n-1}\phi(t-mT)+i\sum_{m=0}^{n+l-1}\phi(t-mT)]} \rangle &= \langle \exp{[-i\sum_{m=0}^{-l-1}\phi_1(t-mT)]} \rangle \cdot \langle \exp{[-i\sum_{m=0}^{-l-1}\phi_2(t-mT)]} \rangle \\
        &= \exp{[+\frac{l}{2} (\langle \phi_1^2 \rangle + \langle \phi_2^2 \rangle)]}.
    \end{split}
\end{equation}
Therefore, generally the second term in Eq.~\ref{g1} can be written as 
\begin{equation}
    \langle \exp{[-i\sum_{m=0}^{n-1}\phi(t-mT)+i\sum_{m=0}^{n+l-1}\phi(t-mT)]} = \exp{[-\frac{|l|}{2} (\langle \phi_1^2 \rangle + \langle \phi_2^2 \rangle)]}
\end{equation}
Substituting this and Eq.~\ref{amp_corr} back to Eq.~\ref{g1}, we obtain 
\begin{equation}
    \begin{split}
        g^{(1)}(\tau) &= \exp{[-\frac{\Gamma_o |\tau|}{2}-\frac{|l|}{2}(\langle \phi_1^2 \rangle+\langle \phi_2^2 \rangle)]}\\
        &= \exp{[(-\frac{\Gamma_o}{2}-\frac{v}{4L}(\langle \phi_1^2 \rangle+\langle \phi_2^2 \rangle))|\tau|]}\\
        &= e^{-\Gamma_\text{decoh}|\tau|/2},
    \end{split}
\end{equation}
where the decoherence rate is a sum of dissipation rate and pure dephasing rate, i.e. $\Gamma_\text{decoh} = \Gamma_o + \Gamma_\phi$, and the pure dephasing is expressed by $\Gamma_\phi=\frac{v}{2L}(\langle \phi_1^2 \rangle+\langle \phi_2^2 \rangle)$, inversely proportional to the cavity length $L$.
At last, the coherence time is defined as 
\begin{equation}
    \begin{split}
        \tau_\text{coh} &= \int_{-\infty}^{\infty} {(g^{(1)}(\tau))^2 d\tau} = \frac{2}{\Gamma_\text{decoh}}.
    \end{split}
\end{equation}
Realize that when there is no pure dephasing ($\Gamma_\phi=0$), the decoherence rate is equal to the dissipation rate ($\Gamma_\text{decoh}=\Gamma_o$), while the coherence time is twice the lifetime ($\tau_\text{coh}=2\tau_o$), same as $T_2=2T_1$ for a two-level system.

\section{Phase noise and power spectral density}
Generally speaking, phonon coherence is determined by both energy dissipation and dephasing in an open system. In this session, we will model the power spectral density (PSD) of the phonon field regarding the existence of both mechanisms. Consider a phonon field $b(t)=\Theta(t)\exp(-\Gamma_ot/2-i\Omega_o t-i\phi(t))$, where $\Theta(t)$ is the Heaviside step function, $\Omega_o$ is its oscillating frequency and $\phi(t)$ is a random variable that models the phase noise. Then its intensity can be expressed as $I(t)=<b^\dagger (t) b(t)>=e^{-\Gamma_o t}$, where $\tau_o=1/\Gamma_o$ is the lifetime. We calculate the degree of first-order temporal coherence 
\begin{equation} \label{g1_1}
\begin{split}
g^{(1)}(\tau) & = \frac{\langle b^\dagger (t) b(t+\tau)\rangle}{\langle b^\dagger (t) b(t) \rangle} \\
& = \frac{\int_{\max\{0,-\tau\}}^{+\infty}{e^{-\Gamma_o t/2+i\Omega_o t+i\phi(t)}\cdot e^{-\Gamma_o (t+\tau)/2-i\Omega_o (t+\tau)-i\phi(t+\tau)} dt}}{\int_0^{+\infty}{e^{-\Gamma_o t}}dt} \\
& = e^{-\Gamma_o |\tau|/2-i\Omega_o \tau} \langle e^{-i\Delta \phi(\tau)}\rangle
\end{split}
\end{equation}
where $\Delta \phi(\tau)=\phi(t+\tau)-\phi(t)$.
Assume $\Delta \phi(\tau)$ represents stationary Gaussian noise, i.e. $P(\Delta \phi(\tau))=\frac{1}{\sqrt{2\pi}\sigma}e^{-[\Delta\phi(\tau)]^2/2\sigma^2}$, where the variance $\sigma^2=\langle [\Delta\phi(\tau)]^2\rangle$. Therefore, 
\begin{equation} \label{Gaussian_noise}
\begin{split}
\langle e^{-i\Delta \phi(\tau)}\rangle & = \int_{-\pi}^{+\pi}{ e^{-i\Delta \phi(\tau)} P(\Delta \phi(\tau))}\cdot d(\Delta \phi(\tau)) \\
& = \frac{1}{\sqrt{2\pi}\sigma} \int_{-\infty}^{+\infty} e^{-i\Delta \phi(\tau)} e^{-[\Delta\phi(\tau)]^2/2\sigma^2}\cdot d(\Delta \phi(\tau)) \\
& = e^{-\sigma^2/2} \\
& = e^{-\langle[\Delta \phi(\tau)]^2\rangle/2}.
\end{split}
\end{equation}
We can then rewrite $g^{(1)}(\tau)$ as 
\begin{equation} \label{g1_2}
g^{(1)}(\tau) = e^{-\Gamma_o |\tau|/2-i\Omega_o \tau} e^{-\langle[\Delta \phi(\tau)]^2\rangle/2}.
\end{equation}
Then, according to the Wiener–Khinchin theorem, the one-sided PSD of the phonon field can be expressed as 
\begin{equation} \label{phonon_psd1}
\begin{split}
S_{bb}[\Omega] & = 2\int_{-\infty}^{+\infty}{g^{(1)}(\tau)\cos{\Omega\tau}\cdot d\tau} \\
& = 2\int_{-\infty}^{+\infty}{d\tau\cdot \cos{\Omega\tau} \cdot e^{-\Gamma_o |\tau|/2-i\Omega_o \tau} e^{-\langle[\Delta \phi(\tau)]^2\rangle/2}} \\
& = 4\int_{0}^{+\infty}{d\tau\cdot \cos{\Omega\tau}\cos{\Omega_o\tau} \cdot e^{-\Gamma_o |\tau|/2} e^{-\langle[\Delta \phi(\tau)]^2\rangle/2}}.
\end{split}
\end{equation}
On the other hand, by definition, $\langle[\Delta \phi(\tau)]^2\rangle=\langle[\phi(t+\tau)-\phi(t)]^2\rangle=2[\langle\phi^2(t)\rangle-\langle\phi(t)\phi(t+\tau)\rangle]$.
Realize that $\langle\phi(t)\phi(t+\tau)\rangle$ is the auto-correlation function o f $\phi(t)$. According to the Wiener-Khinchin theorem, it also forms a Fourier pair with the one-sided PSD of the phase function $S_{\phi\phi}[\Omega]$ as 
\begin{equation} \label{phase_noise_psd}
\langle\phi(t)\phi(t+\tau)\rangle = \frac{1}{2\pi}\int_{0}^{+\infty}{S_{\phi\phi}[\Omega]\cos{\Omega\tau}\cdot d\Omega}
\end{equation}
Thus, 
\begin{equation} \label{phase_coherence}
\begin{split}
\langle[\Delta \phi(\tau)]^2\rangle & = \frac{1}{\pi}\int_{0}^{+\infty}{S_{\phi\phi}[\Omega](1-\cos{\Omega\tau})\cdot d\Omega} \\
& = \frac{2}{\pi}\int_{0}^{+\infty}{S_{\Omega\Omega}[\Omega] (\frac{\sin^2{\Omega\tau/2}}{\Omega^2} )\cdot d\Omega}.
\end{split}
\end{equation}
Substituting it back to eq.(\ref{phonon_psd1}), we obtain
\begin{equation} \label{phonon_psd2}
\begin{split}
S_{bb}[\Omega] & = 4\int_{0}^{+\infty}{d\tau\cdot \cos{\Omega\tau}\cos{\Omega_o\tau} \cdot e^{-\Gamma_o |\tau|/2} e^{-\langle[\Delta \phi(\tau)]^2\rangle/2}} \\
& = 4\int_{0}^{+\infty}{d\tau\cdot e^{-\Gamma_o |\tau|/2}\cos{\Omega\tau}\cos{\Omega_o\tau} \cdot \exp\{ -\frac{1}{\pi}\int_{0}^{+\infty}{S_{\Omega\Omega}[\Omega']\cdot \frac{\sin^2{\Omega'\tau/2}}{\Omega'^2} \cdot d\Omega'} \}}.
\end{split}
\end{equation}
The last step uses the relation $S_{\phi\phi}[\Omega]=S_{\Omega\Omega}[\Omega]/\Omega^2$ as $\Omega(t)=\dot{\phi}(t)$. 

Now let's take the example of white frequency noise, i.e. $S_{\Omega\Omega}[\Omega]=\gamma$ a constant.
Therefore, the PSD of the phonon field can be calculated as 
\begin{equation} \label{phonon_psd3}
\begin{split}
S_{bb}[\Omega] & = 4\int_{0}^{+\infty}{d\tau\cdot e^{-\Gamma_o |\tau|/2}\cos{\Omega\tau}\cos{\Omega_o\tau} \cdot \exp\{ -\frac{1}{\pi}\int_{0}^{+\infty}{S_{\Omega\Omega}[\Omega']\cdot \frac{\sin^2{\Omega'\tau/2}}{\Omega'^2} \cdot d\Omega'} \}} \\
& = 4\int_{0}^{+\infty}{d\tau\cdot e^{-\Gamma_o |\tau|/2}\cdot \cos{\Omega\tau}\cos{\Omega_o\tau} }\cdot e^{-\gamma |\tau|/4} \\
& = 2\int_{0}^{+\infty}{d\tau\cdot e^{-\Gamma |\tau|/2}\cdot [\cos{(\Omega-\Omega_o)\tau}+\cos{(\Omega+\Omega_o)\tau}] } \\
& = \frac{1}{\Gamma/2-i(\Omega-\Omega_o)} + \frac{1}{\Gamma/2+i(\Omega-\Omega_o)} \\
& = \frac{\Gamma}{(\Gamma/2)^2+(\Omega-\Omega_o)^2}
\end{split}
\end{equation}
where $\Gamma=\Gamma_o+\gamma/2$ describing the decoherence rate, and we neglect the negative frequency component since it's one-sided PSD.
We realize that under the assumption of white Gaussian frequency noise, the PSD of the phonon field is Lorentzian and its linewidth is the sum of energy dissipation rate and a dephasing term, i.e. $\Delta\Omega=\Gamma=\Gamma_o+\Gamma_\phi$. Here $\Gamma_\phi=\gamma/2$ is half of the one-sided PSD of the phase noise. 


\section{Phase matching condition} \label{SI_phasematch}
In this section, I will walk you through the theory behind the CABS spectroscopy. This is essentially a three-mode interaction problem, one phonon mode and two optical modes. Note that the phonon modes are discrete modes because they are bounded by the substrate surfaces, while the two optical modes are continuous. 
Let's start with the field quantization.

\subsection{Quantization of the acoustic field}
Following \cite{sipe2016hamiltonian}, we introduce vector field variables $\mathbf{u}(\mathbf{r},t)$ describing the displacement and $\mathbf{\pi}(\mathbf{r},t)$ as their conjugate momenta. Since $\mathbf{u}(\mathbf{r},t)$ and $\mathbf{\pi}(\mathbf{r},t)$ form a canonical pair, they should satisfy the following commutation relations
\begin{equation}\left[u^n(\mathbf{r}),\pi^m(\mathbf{r'})\right]=i\hbar\delta_{nm}\delta(\mathbf{r}-\mathbf{r'}) \quad (m,n=1,2,3),
\end{equation} \label{real_comm}
where $\delta_{nm}$ and $\delta(\mathbf{r}-\mathbf{r'})$ are the Kronecker delta and Dirac delta functions respectively. 
The acoustic Hamiltonian can be formulated as 
\begin{equation} \label{hamilt}
    H^A = \int \frac{\pi^i(\mathbf{r}) \pi^i(\mathbf{r})}{2 \rho(\mathbf{r})} \, d\mathbf{r} + \frac{1}{2} \int S^{ij}(\mathbf{r}) c^{ijkl}(\mathbf{r}) S^{kl}(\mathbf{r}) \, d\mathbf{r},
\end{equation}
where $\rho(\mathbf{r})$ is the density, $c^{ijkl}(\mathbf{r})$ is the stiffness tensor, and
\begin{equation}
    S^{ij}(\mathbf{r}) = \frac{1}{2} \left( \frac{\partial u^i(\mathbf{r})}{\partial r^j} + \frac{\partial u^j(\mathbf{r})}{\partial r^i} \right),
\end{equation}
is the strain tensor. 
Based on this Hamiltonian, we can quickly write down the Heisenberg equations of motion, 
\begin{align} 
\frac{\partial}{\partial t} u^n(\mathbf{r}, t) &= \frac{1}{i\hbar}\left[ u^n(\mathbf{r}), H^A \right]=
\frac{\pi^n(\mathbf{r}, t)}{\rho(\mathbf{r})}, \label{pi} \\
\frac{\partial}{\partial t} \pi^n(\mathbf{r}, t) &= \frac{1}{i\hbar}\left[ \pi^n(\mathbf{r}), H^A \right] = \frac{\partial}{\partial r^j} \left( c^{njkl}(\mathbf{r}) S^{kl}(\mathbf{r}) \right).
\end{align}
From these equations of motion, we can recover the Christoffel equation
\begin{equation}
    \rho(\mathbf{r}) \frac{\partial^2 u^n(\mathbf{r}, t)}{\partial t^2} = \frac{\partial}{\partial r^j} \left( c^{njkl}(\mathbf{r}) S^{kl}(\mathbf{r}) \right).
\end{equation}

Similar to the quantization of electromagnetic fields bounded in an FP cavity, we write down 
\begin{equation} \label{quant}
    \mathbf{u}(\mathbf{r}) = \sum_{m} \sqrt{\frac{\hbar \Omega_m}{2}} \, \mathbf{U}_m(\mathbf{r})\hat{b}_{m} + \text{h.c.},
\end{equation}
and claim that it will be the correct quantitation of the phonon field if the following normalization condition holds,
\begin{equation} \label{norm}
    \int \rho(\mathbf{r}) \, \Omega_m^2 \, \mathbf{U}_m^*(\mathbf{r}) \, \mathbf{U}_{m'}(\mathbf{r}) \, d\mathbf{r} = \delta_{mm'}.
\end{equation}
To confirm it, we substitute Eq.~\ref{quant} back to the Hamiltonian equation~\ref{hamilt} and hope to see the Hamiltonian of a harmonic oscillator.
From Eq.~\ref{pi} we get
\begin{equation}
    \mathbf{\pi}(\mathbf{r}, t) = \rho(\mathbf{r}) \frac{\partial \mathbf{u}(\mathbf{r}, t)}{\partial t} = \sum_m \left( -i \Omega_m \rho(\mathbf{r}) \right) \sqrt{\frac{\hbar \Omega_m}{2}} \, \mathbf{U}_m(\mathbf{r}) \, \hat{b}_m + \text{h.c.}.
\end{equation}
Then substitute it back to the Hamiltonian equation~\ref{hamilt} and we obtain
\begin{equation}
    \begin{split}
    H^A &= \int \frac{d\mathbf{r}}{\rho(\mathbf{r})} 
    \left[ \sum_m (-i \Omega_m \rho(\mathbf{r})) \sqrt{\frac{\hbar \Omega_m}{2}} 
    \mathbf{U}_m(\mathbf{r}) \hat{b}_m + \text{h.c.} \right]^2 \\[10pt]
    &= \sum_{m,m'} \frac{\hbar}{2} \sqrt{\Omega_m \Omega_{m'}} 
    \int \rho(\mathbf{r}) \Omega_m \Omega_{m'} \mathbf{U}_m^*(\mathbf{r}) \mathbf{U}_{m'}(\mathbf{r}) d\mathbf{r} \cdot
    \hat{b}_m^\dagger \hat{b}_{m'} + \text{h.c.} \\[10pt]
    &= \frac{1}{2} \sum_m \hbar \Omega_m 
    \left( \hat{b}_m^\dagger \hat{b}_m + \hat{b}_m\hat{b}_m^\dagger \right),
    \end{split}
\end{equation}
One can also easily confirm that the following commutation relation between $\hat{b}_m$ and $\hat{b}_m^\dagger$ holds, 
\begin{equation}
\left[\hat{b}_m,\hat{b}_{m'}^\dagger\right]=\delta_{m,m'}.
\end{equation}

For simplicity, consider a longitudinal acoustic wave propagating within the \textmu HBAR of length $L$ that has an effective area of $A_a$. It gets reflected at the boundary and forms a standing wave.
Then the scalar field of the standing wave can be expressed by 
\begin{equation}
    U_m(\mathbf{r})=U_m\cdot \cos(q_mz),
\end{equation}
where $q_m=\Omega_m/v=2\pi m/\lambda_\text{ph}$ is the $k$-vector of the $m$-th phonon mode.
The normalization equation~\ref{norm} requires 
\begin{equation}
\begin{split}
    \rho\Omega_m^2\int_V |U_m(\mathbf{r})|^2d\mathbf{r} 
    &= \rho\Omega_m^2\int_{A_a} |U_m|^2 dxdy \cdot \int_0^L \cos^2(q_mz)dz \\
    &= \frac{\rho\Omega_m^2L}{2} |U_m|^2A_a\\
    &= 1,
\end{split}
\end{equation}
which gives 
\begin{equation} \label{A_quant}
    U_m(\mathbf{r}) = \sqrt{\frac{2}{\rho\Omega_m^2 A_aL}}\cdot \cos(q_mz).
\end{equation}

\subsection{Quantization of the optical field}
The quantization of the optical field is similar to that of the acoustic field. The tricky part is that although they form standing waves due to the mirror reflection, they are continuous modes. As you will see in this section, it makes the normalization condition a bit different. The optical Hamiltonian can be written as 
\begin{equation} \label{H_EM}
    H^{\text{EM}} = \frac{1}{2\mu_0} \int B^i(\mathbf{r}) B^i(\mathbf{r}) \, d\mathbf{r} + \frac{1}{2\epsilon_0} \int D^i(\mathbf{r}) \beta_\text{ref}(\mathbf{r}) D^i(\mathbf{r}) \, d\mathbf{r},
\end{equation}
where $\mathbf{B}(\mathbf{r})$ is the magnetic field, $\mathbf{D}(\mathbf{r})$ is the electric displacement, and $\beta_{\text{ref}}(\mathbf{r})=1/\epsilon_{\text{ref}}(\mathbf{r})$ is the inverse of the background permittivity, neglecting any acoustic effects. 

Now since the optical field is continuous, we expand the electric displacement as 
\begin{equation} \label{D_expand}
\mathbf{D}(\mathbf{r}, t) = \int dk \, \sqrt{\frac{\hbar \omega_k}{2}} \, \mathbf{U}_k(\mathbf{r}) \, \hat{a}_k(t) + \text{h.c.},
\end{equation}
with the normalization condition
\begin{equation} \label{EM_norm}
\frac{1}{\epsilon_0} \int d\mathbf{r} \, \beta_\text{ref}(\mathbf{r}) \, \mathbf{U}_k^*(\mathbf{r}) \, \mathbf{U}_{k'}(\mathbf{r}) = \delta(k - k').
\end{equation}
Substituting Eq.~\ref{D_expand} back to Eq.~\ref{H_EM}, we can verify that it is the correct quantization form of the electromagnetic field. 
\begin{equation}
    \begin{split}
        H^{\text{EM}} &= \frac{1}{\epsilon_0} \int D^i(\mathbf{r}) \beta_\text{ref}(\mathbf{r}) D^i(\mathbf{r}) \, d\mathbf{r}\\[10pt]
        &= \frac{1}{\epsilon_0} \int d\mathbf{r} \beta_\text{ref}(\mathbf{r})\left[ \int dk \sqrt{\frac{\hbar \omega_k}{2}} \mathbf{U}_k(\mathbf{r}) \hat{a}_k(t) + \text{h.c.} \right]^2 \\[10pt]
        &= \iint dk \, dk' \, \frac{\hbar}{2} \sqrt{\omega_k \omega_{k'}} \cdot \frac{1}{\epsilon_0} \int d\mathbf{r} \, \beta(\mathbf{r}) \mathbf{U}_k^*(\mathbf{r}) \mathbf{U}_{k'}(\mathbf{r})\cdot \hat{a}_k^\dagger \hat{a}_{k'} + \text{h.c.}, \\[10pt]
        &= \frac{1}{2} \int \hbar \omega_k \left( \hat{a}_k^\dagger \hat{a}_k + \hat{a}_k \hat{a}_k^\dagger \right).
    \end{split}
\end{equation}

For simplicity, we consider a linearly polarized optical field propagating in a medium that has uniform permittivity. Write down the optical field as 
\begin{equation}
    U_k(\mathbf{r})=U_k\cdot \sin(kz) \quad (k>0).
\end{equation}
Then the normalization condition Eq.~\ref{EM_norm} requires
\begin{equation}
    \begin{split}
    \frac{1}{\epsilon_0} \int d\mathbf{r} \, \beta(\mathbf{r}) \, U_k^*(\mathbf{r}) \, U_{k'}(\mathbf{r}) 
    &= \frac{\beta_\text{ref}}{\epsilon_0} \int_{A_o} U_k^*U_{k'} dxdy \cdot \int_{0}^{\infty} \sin(kz) \sin(k' z) \, dz\\[10pt]
    &= \frac{1}{\epsilon_0\epsilon_\text{ref}}|U_k|^2 A_o \cdot \frac{\pi}{2}\delta(k-k')\\[10pt]
    &= \delta(k-k').
    \end{split}
\end{equation}
It gives 
\begin{equation} \label{EM_quant}
    U_k(\mathbf{r}) = \sqrt{\frac{2\epsilon_0\epsilon_\text{ref}}{\pi A_o}}\cdot \sin(kz),
\end{equation}
where $A_o$ is the effective area of the optical modes.

\subsection{The zero-point coupling rate}
\begin{figure}
    \centering
    \includegraphics[width=0.8\linewidth]{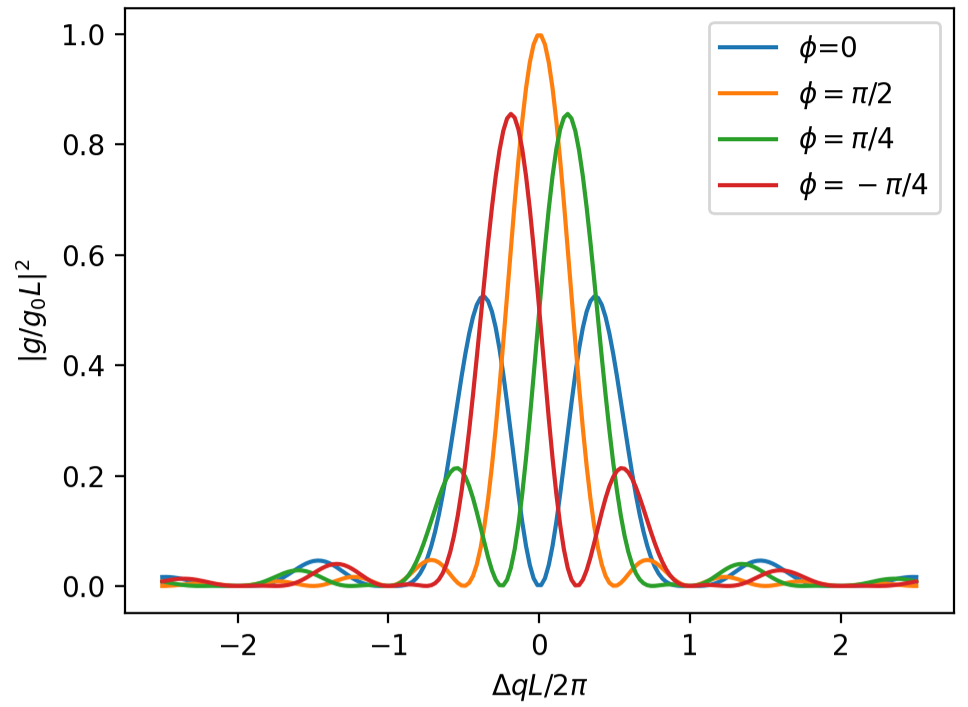}
    \caption{Phase matching profile. $L$ is the substrate thickness and $\Delta q = k_p + k_S - q_m$ is the $k$ mismatch. It is largely modulated by the phase delay $\phi=(k_p+k_S)d\approx 2\pi d/\lambda_\text{ph}$ caused by the spacing between the optical mirror and substrate.}
    \label{fig:Ch3_gprofile}
\end{figure}
Now that we have quantized the optical field and acoustic field, neglecting the moving boundary effect (this is usually infinitesimal in bulk devices like our \textmu HBARs), let's write down the optomechanical interaction Hamiltonian only addressing the photoelastic effect,
\begin{equation} \label{H_int}
\begin{split}
    H^\text{int} &= \frac{1}{2\epsilon_0} \int D^i(\mathbf{r}) D^j(\mathbf{r}) p^{ijlm} S^{lm}(\mathbf{r}) \\[10pt]
    &= \frac{1}{2\epsilon_0} \int D^i(\mathbf{r}) D^j(\mathbf{r}) p^{ijlm}  \frac{\partial u^l(\mathbf{r})}{\partial r^m},
\end{split}
\end{equation}
where $p^{ijkl}$ is the photoelastic tensor.

Consider two laser beams, one pump beam $(k_p, \omega_p)$ and one Stokes beam $(k_S, \omega_S)$, are interacting with the acoustic wave $(q_m, \Omega_m)$. Assume both pump and Stokes beams are x-polarized and the acoustic wave is a longitudinal wave propagating along z-direction. Substituting Eq.~\ref{A_quant} and Eq.~\ref{EM_quant} into Eq.~\ref{H_int}, we get 
\begin{align}
        H^\text{int} 
        &= \frac{1}{2\epsilon_0} \int_V D^2(\mathbf{r}) p^{13} \partial_z u(\mathbf{r})\notag\\[10pt]
        &= \frac{p^{13}}{2\epsilon_0} \int_V d\mathbf{r} \left[ \int dk_p \sqrt{\frac{\hbar \omega_{k_p}}{2}} U_{k_p} \sin(k_p z) \hat{a}_{k_p} + \int dk_S \sqrt{\frac{\hbar \omega_{k_S}}{2}} U_{k_S} \sin(k_S z) \hat{a}_{k_S} + \text{h.c.} \right]^2 \notag\\[10pt]
        & \hspace{20mm}\cdot \left[ \sum_m \sqrt{ \frac{\hbar \Omega_m}{2}} U_m (-q_m) \sin(q_m(z - d)) \hat{b}_m + \text{h.c.} \right] \notag\\[10pt]
        & = \frac{p^{13}}{2\epsilon_0} \sum_m \sqrt{\frac{\hbar^3 \omega_{k_p} \omega_{k_S} \Omega_m}{8}} \iint dk_p dk_S U_{k_p} U_{k_S} (-q_m) U_m A_{ao}\cdot \notag\\[10pt]
        & \hspace{20mm}\int_d^{d+L} dz \sin(k_p z) \sin(k_S z) \sin(q_m (z - d)) \cdot 2(\hat{a}_{k_p}^\dagger \hat{a}_{k_S} \hat{b}_m + \text{h.c.}) \notag\\[10pt]
        & = -\sum_m \iint dk_p dk_S \int_d^{d+L} \sin(\Delta q z + q_m d)dz \cdot(\hat{a}_{k_p}^\dagger \hat{a}_{k_S} \hat{b}_m + \text{h.c.})\cdot \notag\\[10pt]
        & \hspace{20mm} \frac{q_m p^{13}}{4\epsilon_0} \sqrt{\frac{\hbar^3 \omega_{k_p} \omega_{k_S} \Omega_m}{8}} U_{k_p} U_{k_S} U_m A_{ao}.
\end{align}
Here the optical mirror surface is defined as $z=0$ and $d$ is the spacing between the mirror surface and the substrate surface. The substrate has thickness $L$, thus the interaction Hamiltonian integrates from $d$ to $d+L$. $\Delta q = k_p + k_S - q_m$ is the phase mismatch.
Here I assume the effective areas of the laser beams and acoustic wave are the same, i.e. $A_a=A_o=A_{ao}$. I also used the identity 
\begin{equation}
    \sin(a) \sin(b) \sin(c) = \frac{1}{4} \left( \sin(c + a - b) + \sin(c - a + b) - \sin(c + a + b) - \sin(c - a - b) \right),
\end{equation}
in the derivation. 
Comparing this with 
\begin{equation}
    H^\text{int} = -\sum_m \iint \frac{dk_p dk_S}{2\pi} \hbar g(k_p,k_S,g_m) \cdot(\hat{a}_{k_p}^\dagger \hat{a}_{k_S} \hat{b}_m + \text{h.c.}),
\end{equation}
we get the zero-point coupling rate of the three-wave interaction
\begin{equation} \label{g}
    \begin{split}
        g(k_p,k_S,g_m) &= \frac{\pi p^{13} q_m}{4\epsilon_0}\sqrt{\frac{\hbar}{2}\omega_{k_p}\omega_{k_S}\Omega_m}\cdot U_{k_p}U_{k_S}U_mA_{ao}\\[10pt]
        & \hspace{20mm} \cdot \int_d^{d+L} \sin(\Delta q z + q_m d)dz\\[10pt]
        &= \frac{\epsilon_\text{ref} \  p^{13}q_m}{2}\sqrt{\frac{\hbar \omega_{k_p}\omega_{k_S}}{\rho \Omega_m A_{ao} L}}\cdot \int_d^{d+L} \sin(\Delta q z + q_m d)dz\\[10pt]
        &= \frac{p^{13} n^3 \omega^2}{c}\sqrt{\frac{\hbar}{\rho \Omega_m A_{ao} L}} \cdot \int_d^{d+L} \sin(\Delta q z + q_m d)dz\\[10pt]
        &= g_0\cdot \int_d^{d+L} \sin(\Delta q z + q_m d)dz.
    \end{split}
\end{equation}
Note that $g(k_p,k_S,g_m)$ has unit of [Hz$\cdot$m], and $g_0$ has unit of [Hz], because
\begin{equation}
    \begin{split}
        \int_d^{d+L} \sin(\Delta q z + q_m d)dz &= \int_0^L \sin(\Delta q z + \phi)dz\\[10pt]
        &= \frac{1}{\Delta q}\left[ \cos\phi-\cos(\Delta q L+\phi) \right]\\[10pt]
        &= \frac{2}{\Delta q}\sin(\Delta q L/2)\sin(\phi+\Delta q L/2)\\[10pt]
        &= L\cdot \frac{\sin(\Delta q L/2)\sin(\phi+\Delta q L/2)}{\Delta q L/2}.
    \end{split}
\end{equation}
Here $\phi=(k_p+k_S)d\approx q_md=2\pi d/\lambda_\text{ph}$ is the phase delay introduced by the spacing between the mirror and substrate and it can largely modulate the phase matching profile. 
When $\phi=0$, 
\begin{equation}
    g(k_p,k_S,g_m)=g_0 L\cdot \sin(\Delta q L/2) \text{sinc}(\Delta q L/2).
\end{equation}
When $\phi=\pi/2$, 
\begin{equation}
    g(k_p,k_S,g_m)=g_0 L \cdot\text{sinc}(\Delta q L).
\end{equation}
Fig.~\ref{fig:Ch3_gprofile} shows the phase matching profiles under different phase delays. It's very sensitive to the phase delay, thus to the spacing $d$ given how small the phonon wavelength is. I'd like to remind you that one FSR means that single-trip phase shifts by $\Delta q L=\pi$, therefore in Fig.~\ref{fig:Ch3_gprofile} it will be $\Delta q L=\pi/2\pi=1/2$. Therefore, you can see at most 2 or 3 longitudinal mode families in the main lobe of the phase matching profile in the spectral measurement. 


\end{document}